\documentclass{article}

\usepackage{epsfig}
\usepackage{geometry}

\def\e{\epsilon}
\def\ve{\varepsilon}
\def\G{\Gamma}
\def\le{\left(}
\def\ri{\right)}
\def\no{\nonumber}
\def\rar{\rightarrow}

\begin{document}

\begin{titlepage}
\flushright{BI-TP 2012/44}
\vskip 2cm
\begin{center}
{\Large \bf  Box ladders  in non-integer dimension} \\
\vskip 1cm  
Ivan Gonzalez $^{(a,b)}$ and Igor Kondrashuk $^{(c,d)}$   
\vskip 5mm  
{\it  (a) Universidad de Valparaiso, Departamento de F\'\i sica y
Astronomia, \\
Avenida Gran Bretana 1111, Valparaiso, Chile} \\
{\it  (b) Universidad T\'ecnica Federico Santa Maria, and Centro Cient\'ifico-Tecnológico de Valparaiso, 
Casilla 110-V, Valparaiso, Chile} \\
{\it  (c) Departamento de Ciencias B\'asicas,  Universidad del B\'\i o-B\'\i o, \\ 
          Campus Fernando May, Casilla 447, Chill\'an, Chile} \\
{\it  (d) Faculty of Physics, University of Bielefeld, D-33501 Bielefeld, Germany} 
\end{center}
\vskip 0.5cm

\begin{abstract}
We construct a family of triangle-ladder diagrams which may be calculated by making use of Belokurov-Usyukina loop reduction 
technique in $d = 4 -2\ve$ dimensions. The main idea of the approach proposed in the present paper consists in generalization of this loop reduction technique 
existing in  $d = 4$ dimensions. The recursive formula relating the result for $L$-loop triangle ladder diagram of this family and the result  
for $(L-1)$-loop triangle ladder diagram of the same family  is derived. Since the method proposed in the present paper combines 
analytic and dimensional regularizations, at the end of the calculation we have to remove the analytic regularization 
by taking the double uniform limit in which the parameters of the analytic regularization are vanishing. In this limit on the left hand side of the 
recursive relations we obtain in the position space the diagram in which the indices of the rungs are 1  and all the other indices are 
$1-\ve.$ Fourier transformation of the diagrams of this type gives the momentum space diagrams which have  indices of the rungs equal to  $1-\ve$  and 
all the other indices 1. Via conformal transformation of the dual space image of this momentum space representation 
we relate such a family of the triangle ladder momentum diagrams to a family of the box ladder momentum diagrams in which the
indices of the rungs are equal to $1-\ve$ and all the other indices are 1. Since any diagram from this family can be reduced to one-loop diagram,
the proposed generalization of the Belokurov-Usyukina loop reduction technique to non-integer number of dimensions allows us to calculate this family of box-ladder diagrams 
in the momentum space explicitly in terms of Appell's hypergeometric function  $F_4$ without expanding in powers of parameter 
$\ve$ in an arbitrary kinematic region in the momentum space.  
\end{abstract}
\end{titlepage}

\section{Introduction}

Box-ladder massless diagrams with all the indices equal to 1 in the momentum space representation (the m.s.r.) played a remarkable role in high energy physics, 
for example their sum represents a particular solution to the Bethe-Salpeter equation \cite{Broadhurst:2010ds}. 
Also, this family of diagrams contributed to BDS anzatz for four-point all-order gluon amplitude  in maximally supersymmetric Yang-Mills theory \cite{Bern:2005iz,Smirnov}. 
The present situations with this family of diagrams are different in the case of $d=4$ and in the case of  $d=4-2\ve.$
In Ref.  \cite{Broadhurst:2010ds} the off-shell result for the box ladder diagrams obtained in  Refs. \cite{Usyukina:1992jd,Usyukina:1993ch}
at any loop order for $d=4$ has been used. However, the off-shell all order in loop result for these diagrams is unknown in $d=4-2\ve$ dimensions. 
The on-shell all order in loop results for the box ladder diagrams  in non-integer dimension are unknown too, while in the BDS anzatz  of Refs. \cite{Bern:2005iz,Smirnov}
the dimensionally regularized amplitudes in $d=4-2\ve,$ in which the on-shell values of these ladder diagrams contribute, 
have been calculated up to the three-loop order. The results of Refs. \cite{Usyukina:1992jd,Usyukina:1993ch} cannot be used to calculate 
the amplitudes since the on-shell infrared divergencies appear which needed to be regularized by dimensional regularization.   
Moreover, at present the known on-shell results for the box-ladder diagrams exist in the form of expansions in terms of $\ve$ up to certain power of it \cite{Smirnov}.  
As to the off-shell results for dimensionally regularized diagrams known in all the orders in $\ve,$ the triangle off-shell massless one-loop diagram is the unique example
that can be calculated for arbitrary indices and is expressed in terms of hypergeometric Appell's function $F_4$   \cite{Boos:1990rg,Davydychev:1992xr}.

Calculation at any loop orders  in $d=4$ for the triangle ladder was based on a recursive procedure found in
Refs. \cite{Belokurov:1983km,Usyukina:1983gj,Usyukina:1991cp}. This trick allowed to the authors of  \cite{Belokurov:1983km,Usyukina:1983gj,Usyukina:1991cp}
to relate the $L$-loop result for the triangle scalar massless ladder to $(L-1)$-loop result for the same ladder. Thus, the triangle ladder with an arbitrary number of rungs can be reduced 
to one-loop triangle diagram that is known explicitly for any indices  \cite{Boos:1990rg,Davydychev:1992xr}. 
The triangle and box ladders in $d=4$ are related by conformal transformation of coordinates in the dual space \cite{Broadhurst:1993ib,Usyukina:1992jd,Usyukina:1993ch,Kondrashuk:2009us},
so that the triangle ladder result is equivalent to the box ladder result. It was the first family of the four-point massless diagrams 
that has been calculated for an arbitrary number of loops. However, the procedure of 
\cite{Belokurov:1983km,Usyukina:1983gj,Usyukina:1991cp} works in $d=4$ dimensions only, this is why the result at any loop order  for the box ladders exists 
in $d=4$ only. In Ref. \cite{Gonzalez:2012gu} 
the formula of \cite{Usyukina:1991cp} for the reduction of the triangle ladder with two rungs to the scalar one-loop triangle (this is the first line 
of Fig. (1) of the paper \cite{Allendes:2012mr}) has been generalized to non-integer 
number of dimensions  $d=4-2\ve.$ That formula of \cite{Usyukina:1991cp} was a key step to more complicated recursive formula
of the loop reduction published in paper \cite{Belokurov:1983km} (third line of Fig. (1) of  paper \cite{Allendes:2012mr}). 
We refer to the third line of Fig. (1) of paper \cite{Allendes:2012mr}  as to the Belokurov-Usyukina loop reduction technique \cite{Gonzalez:2012gu}.

In the present paper the  Belokurov-Usyukina loop reduction technique (third line of Fig. (1) of  paper \cite{Allendes:2012mr})
is generalized to non-integer number of dimensions $d=4-2\ve.$  The idea is simple. We relate the three-rung triangle ladder and the two-rung triangle ladder in such a way 
that it looks like a recursive formula. Each step of the proposed construction repeats the corresponding step of the construction in the integer dimensions 
$d=4,$ described in detail in Ref.  \cite{Allendes:2012mr}. The key starting point is the formula  of Ref. \cite{Gonzalez:2012gu}. Then, we show how to calculate a family of
triangle ladder diagrams in the position space with an arbitrary number of rungs by making use of the generalized Belokurov-Usyukina loop reduction technique.

We must mention that this family of ladder diagrams does not have all the indices $1-\ve$ in the position space. 
The rung indices remain to be 1. We do not know at present how to make the generalization of 
Belokurov-Usyukina loop reduction technique to obtain the triangle ladder with all the indices $1-\ve$ in the position space in $d=4-2\ve.$
After switching to the momentum representation, we obtain  the triangle diagram with index  $1-\ve$ on the rungs of the ladder and with all 
the other indices 1. Any box ladder diagram with index  $1-\ve$ on the rungs and all the other indices 1 in the m.s.r. can be 
obtained from the corresponding triangle ladder diagram via conformal transformation in the dual space. This trick generalizes the 
corresponding trick for the case of $d=4$ \cite{Broadhurst:1993ib,Kondrashuk:2009us}.

\section{Belokurov-Usyukina loop reduction technique in $d=4-2\ve$}

\begin{figure}[h!] 

\centering\includegraphics[scale=.5]{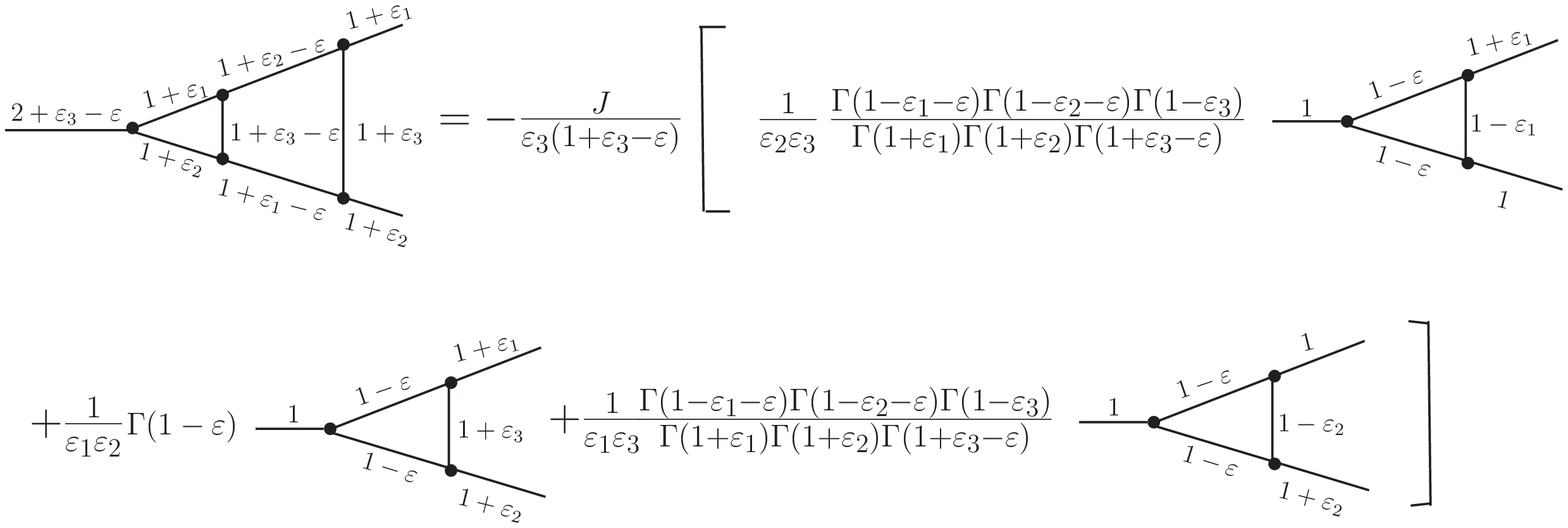}
\vspace{-0.4cm}
 \caption{\footnotesize Formula of Ref. \cite{Gonzalez:2012gu} }
\label{figure01}
 \end{figure}

In Ref.\cite{Gonzalez:2012gu} the formula shown in Fig. (\ref{figure01}) has been derived for non-integer number of dimensions. This has been done by using the 
uniqueness method with applying star-triangle and triangle-star relations \cite{Unique,Vasiliev:1981dg,Vasil,Kazakov:1984bw}. The measure 
redefinition done in Ref.\cite{Cvetic:2006iu} has been applied here too in order to remove powers of $\pi$ on the r.h.s. of the diagrammatic 
relations. Thus, the measure which is used in the internal (bold) vertices of the diagrams is

\begin{eqnarray*}
Dx \equiv \pi^{-\frac{d}{2}}d^d x.
\end{eqnarray*}

The formulae for the chain integration and for the star-triangle integration (uniqueness case) are Eq. (1) and Eq. (2) of Ref. 
\cite{Allendes:2012mr}.

The formula of Fig. (\ref{figure01}) is the key step we start with. 

The factor $J$ of  Ref.\cite{Gonzalez:2012gu} is defined as 

\begin{eqnarray*}
J =   \frac{\G(1-\ve_1)\G(1-\ve_2) \G(1-\ve_3) }{\G(1+\ve_1-\ve)\G(1+\ve_2-\ve) \G(1+\ve_3-\ve) }. 
\end{eqnarray*}

The analytic regularization is applied in combination with the dimensional regularization. The analytic regularization has three parameters which 
satisfy the condition

\begin{eqnarray*}
\ve_1 + \ve_2 + \ve_3 = 0. 
\end{eqnarray*}  

\begin{figure}[h!] 
\centering\includegraphics[scale=.5]{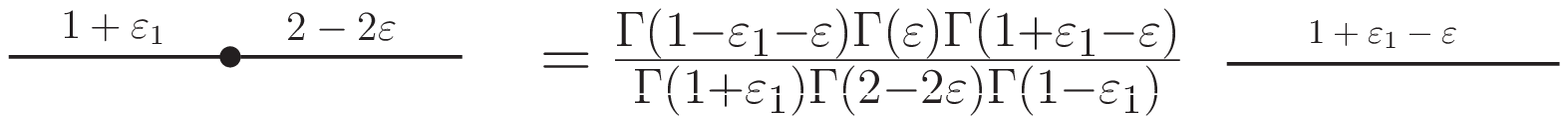}
\vspace{-0.4cm}
 \caption{\footnotesize In order to change the index of the external legs we do this integral convolution }
\label{figure02}
 \end{figure}

To develop the formula of  Fig. (\ref{figure01}) to higher loops, we need to change the indices of the external legs. For this purpose we need to integrate 
external right points of each diagram in Fig. (\ref{figure01}) with the line having index $2-2\e$, as it is shown in  Fig. (\ref{figure02}).

\begin{figure}[h!!!] 
\centering\includegraphics[scale=.5]{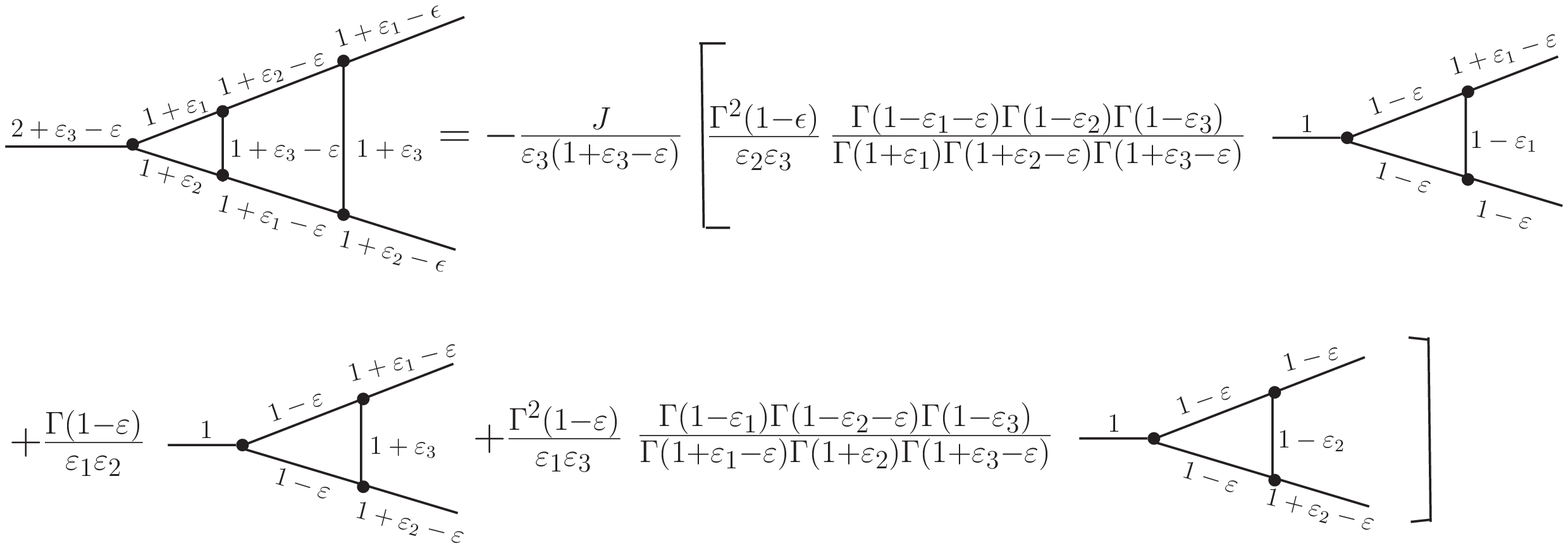}
\vspace{-0.4cm}
 \caption{\footnotesize External points of Fig. (\ref{figure01}) were convoluted with lines having index $2-2\e$ }
\label{figure03}
 \end{figure}

\begin{figure}[h!!!] 
\centering\includegraphics[scale=.5]{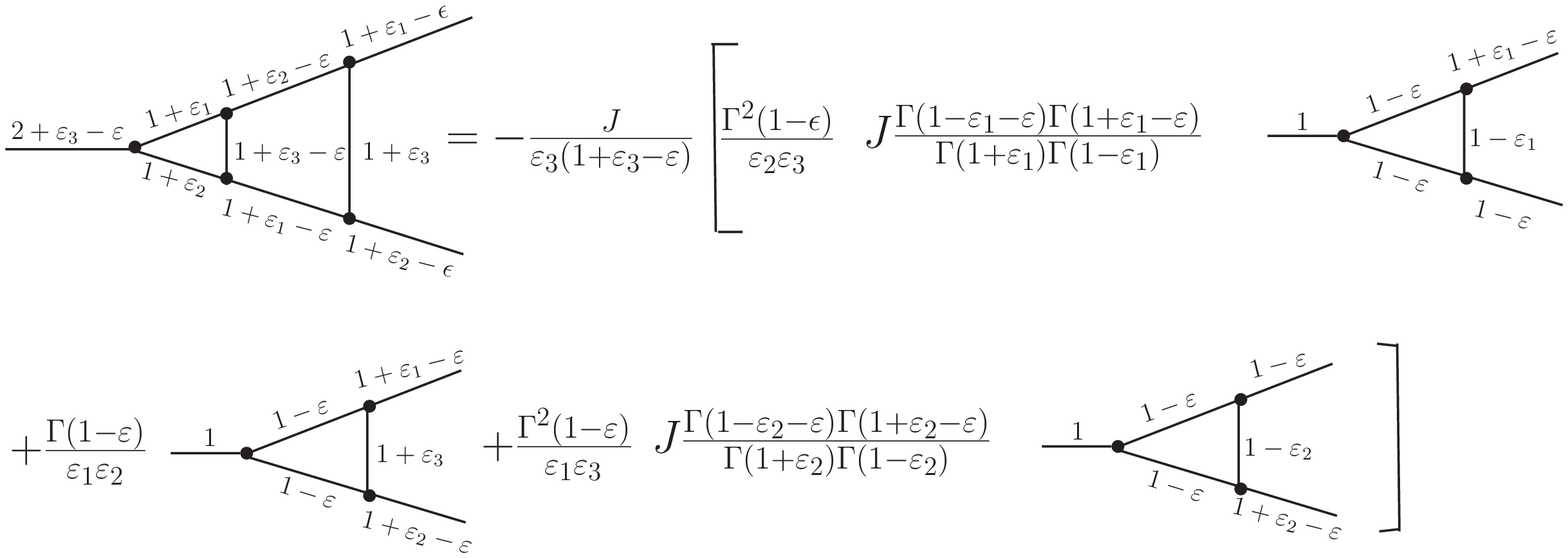}
\vspace{-0.4cm}
 \caption{\footnotesize Integral relation of Fig. (\ref{figure03}) rewritten  with $J$ on the r.h.s. }
\label{figure04}
 \end{figure}

\begin{figure}[h!!!] 
\centering\includegraphics[scale=.5]{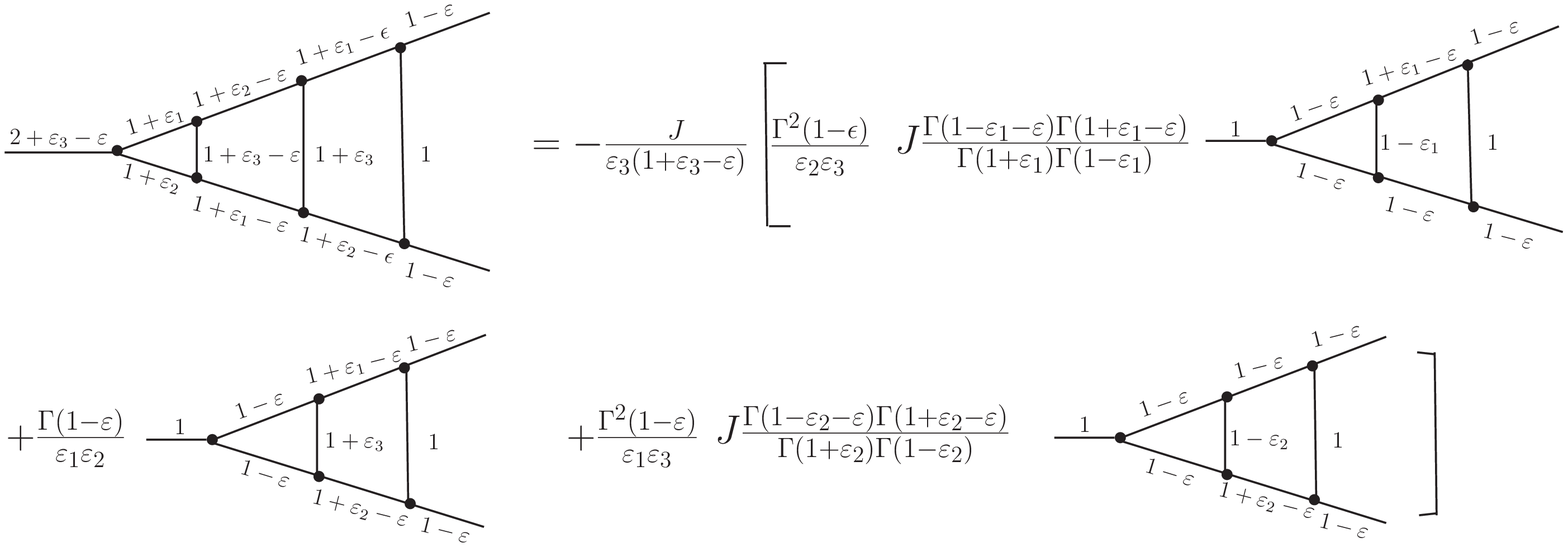}
\vspace{-0.4cm}
 \caption{\footnotesize  This is Fig. (\ref{figure04}) integrated with three new propagators on the r.h.s.    }
\label{figure05}
 \end{figure}

\begin{figure}[h!!!] 
\centering\includegraphics[scale=.5]{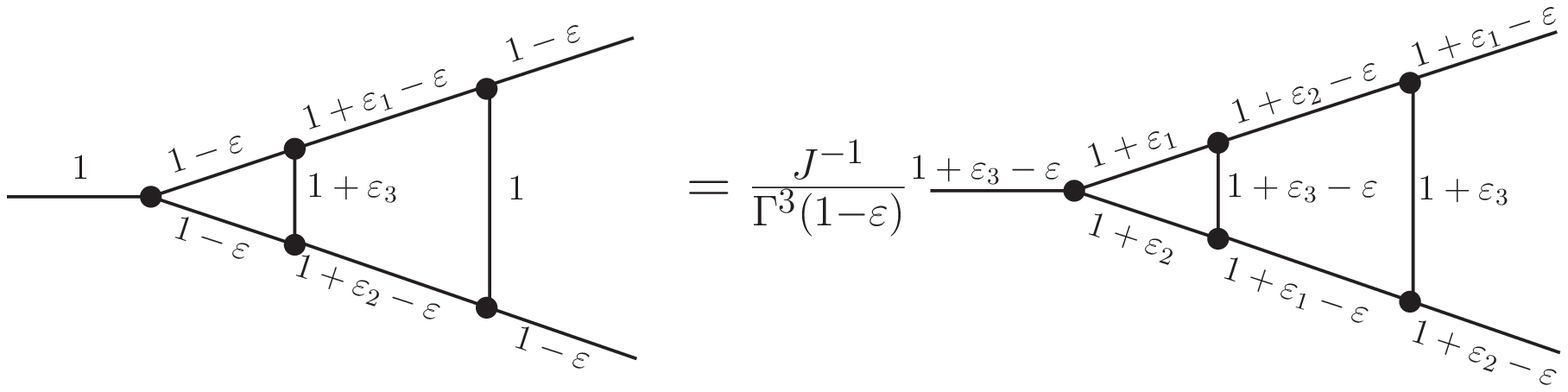}
\vspace{-0.4cm}
 \caption{\footnotesize  Identity for the second diagram in the r.h.s. of  Fig. (\ref{figure05})} 
\label{figure06}
 \end{figure}

\begin{figure}[h!!!] 
\centering\includegraphics[scale=.5]{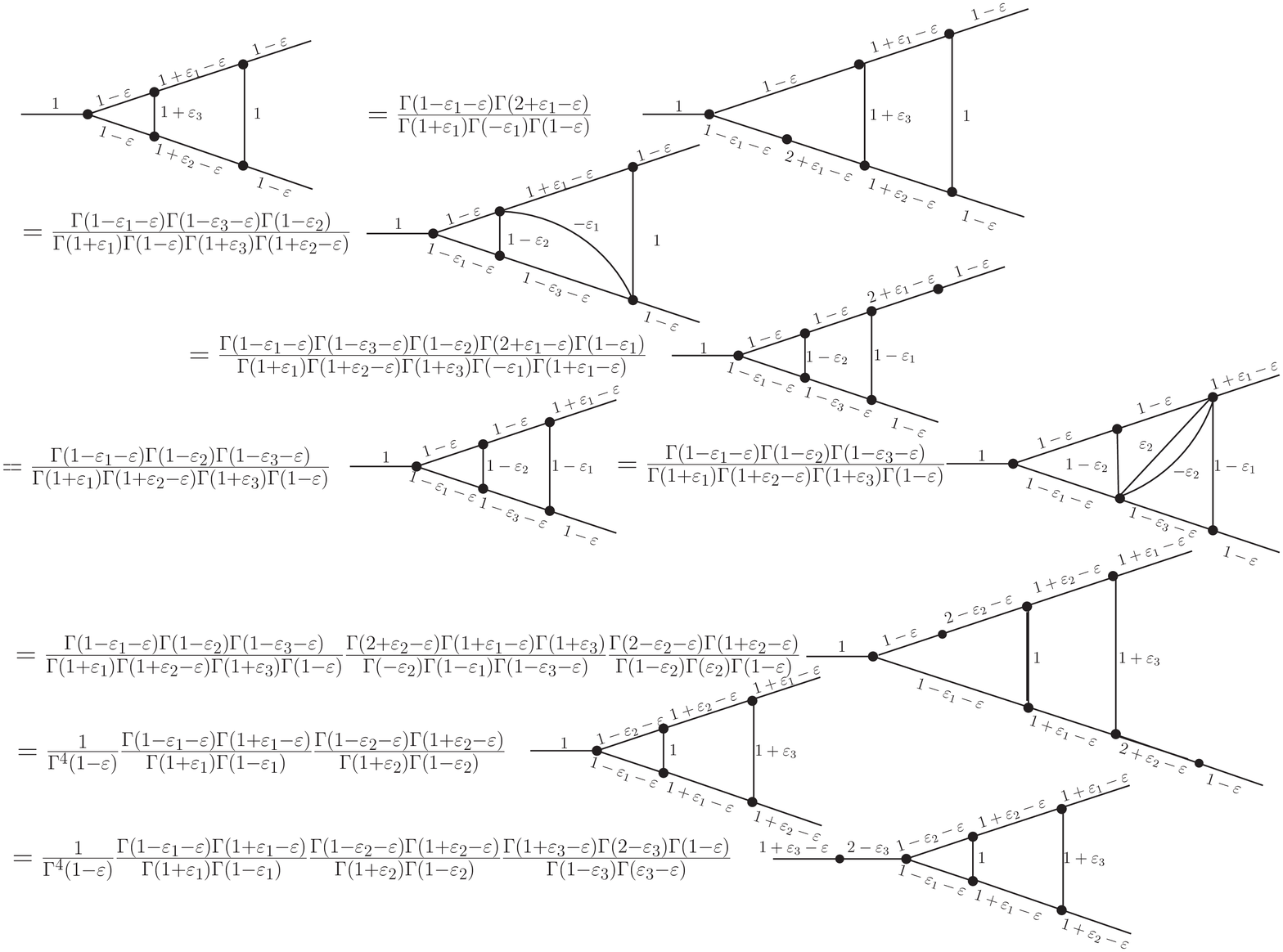}
\vspace{-0.4cm}
 \caption{\footnotesize Proof of the identity depicted in Fig. (\ref{figure06}) }
\label{figure07}
 \end{figure}

\begin{figure}[h!!!] 
\centering\includegraphics[scale=.5]{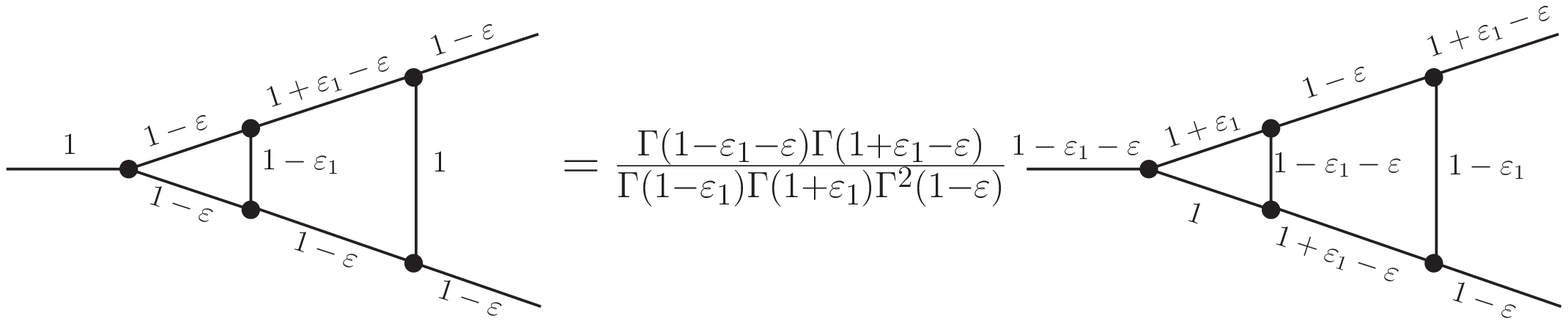}
\vspace{-0.4cm}
 \caption{\footnotesize  Identity for the first diagram in the r.h.s. of  Fig. (\ref{figure05})}
\label{figure08}
 \end{figure}

\begin{figure}[h!!] 
\centering\includegraphics[scale=.5]{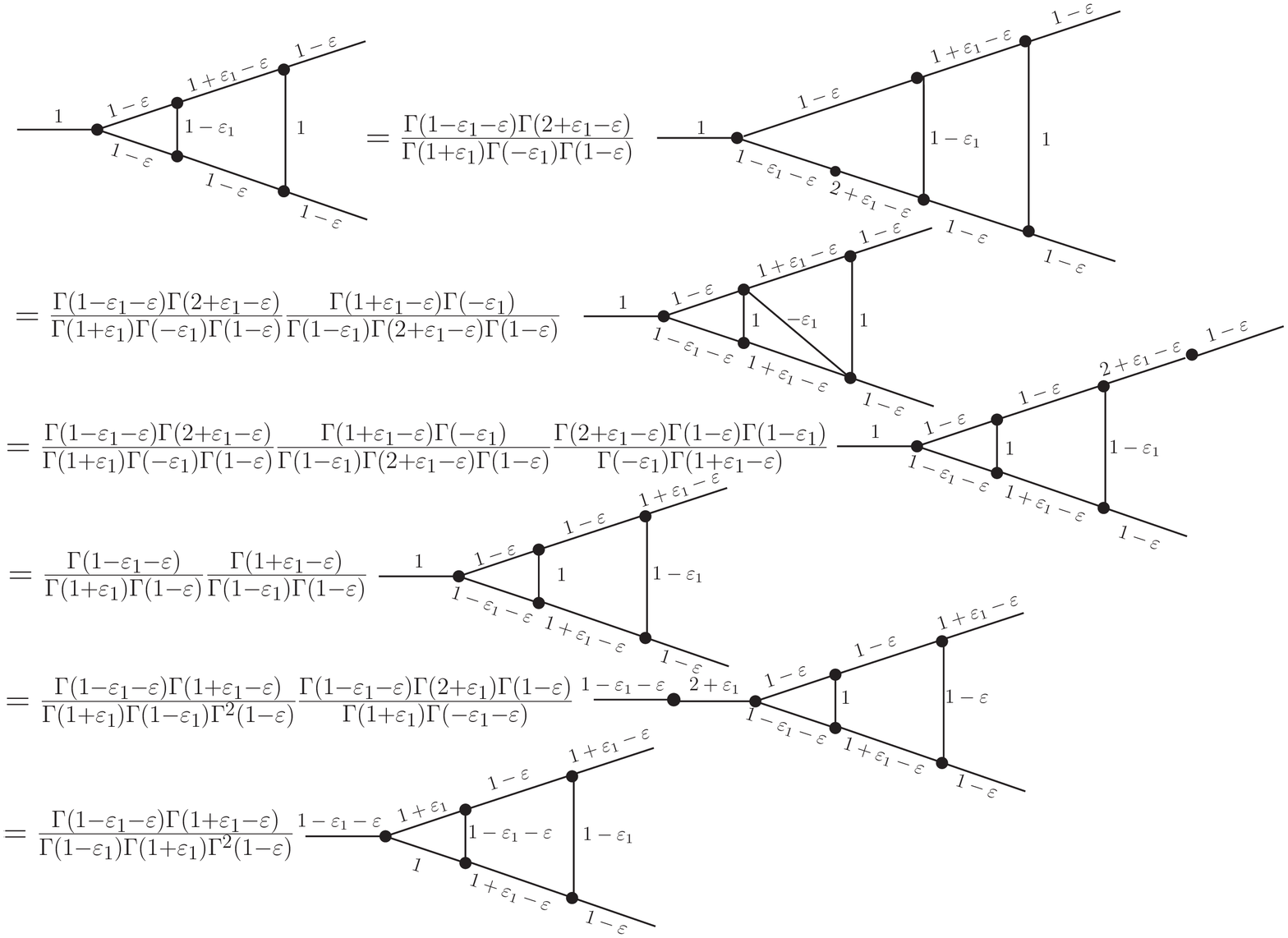}
\vspace{-0.4cm}
 \caption{\footnotesize  Proof of the identity depicted in Fig. (\ref{figure08})} 
\label{figure09}
 \end{figure}

\begin{figure}[h!!!] 
\centering\includegraphics[scale=.5]{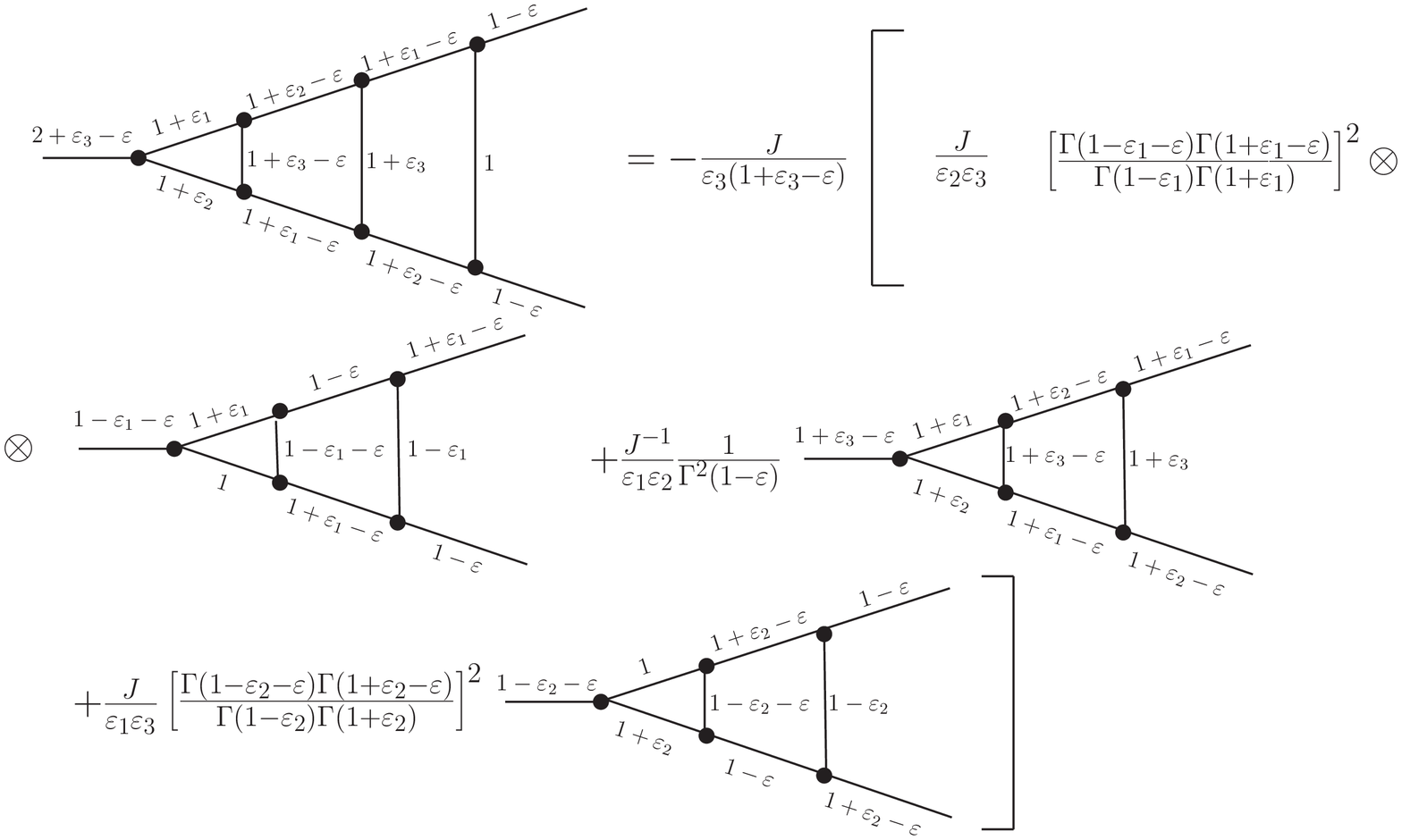}
\vspace{-0.4cm}
 \caption{\footnotesize  Loop reduction in $d= 4-2\ve$ dimensions }
\label{figure10}
 \end{figure}

\begin{figure}[h!!!] 
\centering\includegraphics[scale=.5]{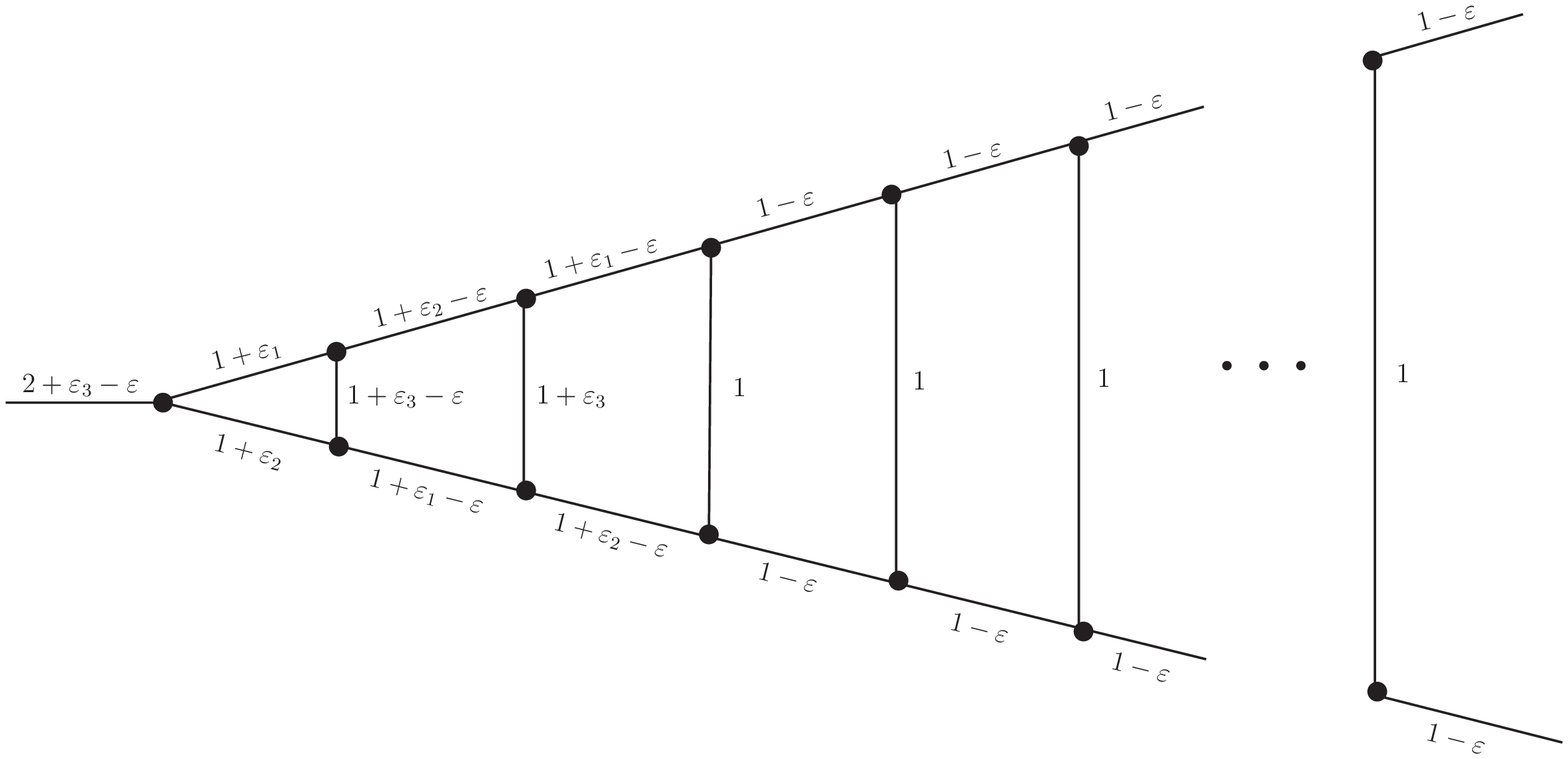}
\vspace{-0.4cm}
 \caption{\footnotesize  Triangle ladder diagrams for which the recursive structure has been found}
\label{figure11}
 \end{figure}

As the result of this convolution, Fig. (\ref{figure01}) transforms to Fig. (\ref{figure03}).
Thus, we have proved the integral relation of Fig. (\ref{figure04}).
As the next step, we integrate the r.h. sides of each diagram in Fig. (\ref{figure04}) with three new propagators. 
The result is shown in Fig. (\ref{figure05}).

Each diagram on the right hand side of the formula depicted in  Fig. (\ref{figure05}) can be transformed to another diagram with a distinct distribution 
of indices by doing a set of simple transformations.
Usually, this can be done by inserting points into the propagators and by using star-triangle and triangle-star relations. 
For example, for the second diagram on the r.h.s. in  Fig. (\ref{figure05}) the identity depicted in Fig. (\ref{figure06}) is valid.
This identity has been proved in Fig. (\ref{figure07}).  Similar transformations with using  the same tricks can be applied to prove the identity for the first 
diagram on the r.h.s. of Fig. (\ref{figure05}) which is depicted in Fig. (\ref{figure08}).
The proof is given in Fig. (\ref{figure09}). As the result of these identities we have found the formula shown in Fig. (\ref{figure10}). 
The formula given in Fig. (\ref{figure10}) has a recursive 
structure since the first and third  diagrams coincide with the second diagram on the r.h.s. of Fig. (\ref{figure10}) when the parameters of the analytic regularization take  
the values $\ve_2 =0 $ or $\ve_1 =0,$ respectively. At the same time, this second diagram on the r.h.s. of Fig. (\ref{figure10}) has the same indices which 
the diagram on the l.h.s. of Fig. (\ref{figure10}) has on its leftmost eight elements.

The recursive formula obtained in Fig. (\ref{figure10}) can be used to represent  
the triangle ladder diagrams of the type depicted in Fig. (\ref{figure11}) in the non-integer number of dimensions $d=4-2\ve$ 
with an arbitrary number of rungs as a linear combination of one-loop triangle diagrams with coefficients that become singular 
in the limit of vanishing $\ve_1$ and $\ve_2.$

\section{Recursive integral relations in momentum space}

In the previous section all the diagrammatic relations  depicted in the figures to which we referred   were understood as the relations between 
the integrals that may be read off from the position space representation (the p.s.r.) of the diagrams. These integral relations in  position space may be transformed to the 
corresponding integral relations in momentum space. To make this transformation we repeat the procedure described in Ref. \cite{Allendes:2012mr}.   
We should replace each factor in the integrand of the p.s.r. with the integral Fourier transform of the corresponding factor 
in the m.s.r. Then, we need to integrate over the coordinates of internal vertices. In such a way we create Dirac $\delta$ functions, corresponding to the 
momentum conservation in each vertex of integration in position space, that is, in the internal vertices. The  momentum integrals over loop momenta 
will be the Fourier transforms of the integrals in position space.

The definition of the integral measure in momentum space is done with the same factor as in position space,  
\begin{eqnarray}\label{k-measure}
Dk \equiv \pi^{-\frac{d}{2}}d^d k,
\end{eqnarray}
to avoid powers of $\pi$ in the corresponding momentum integrals.  The three-point one-loop momentum scalar integral 
\begin{eqnarray*}
J(\nu_1,\nu_2,\nu_3) = \int~Dk~\frac{1}{\left[(k + q_1)^2\right]^{\nu_1} \left[(k + q_2)^2 \right]^{\nu_2}
\left[(k + q_3)^2\right]^{\nu_3}}, 
\end{eqnarray*}
may be re-presented in terms of the MB transform, as it has been done in Ref. \cite{Usyukina:1992jd},  
\begin{eqnarray*}
J(\nu_1,\nu_2,\nu_3) = \frac{1}{\Pi_{i} \G(\nu_i) \G(d-\Sigma_i \nu_i)} \frac{1}{{(p^2_3)}^{ \Sigma \nu_i -d/2}}
\oint_C dz_2~dz_3 x^{z_2}  y^{z_3}
\left\{ \G \le -z_2 \ri \G \le -z_3 \ri \right.\no\\
\left. \G \le -z_2 -\nu_2-\nu_3 + d/2 \ri \G \le -z_3-\nu_1-\nu_3 + d/2 \ri 
\G \le z_2 + z_3  + \nu_3 \ri  \G \le  \Sigma \nu_i - d/2 + z_3 + z_2 \ri \right\} \equiv \label{J-arb}\\
\equiv \frac{1}{{(p^2_3)}^{ \Sigma \nu_i -d/2}}\oint_C dz_2~dz_3 x^{z_2}  y^{z_3} 
D^{(z_2,z_3)} [\nu_1,\nu_2,\nu_3]. \no
\end{eqnarray*}
The definition for the function $D$ is the same as in Ref.\cite{Allendes:2012mr},  
\begin{eqnarray*}
D^{(z_2,z_3)}[\nu_1,\nu_2,\nu_3] = \frac{ \G \le -z_2 \ri \G \le -z_3 \ri \G \le -z_2 -\nu_2-\nu_3 + d/2 \ri 
\G \le -z_3-\nu_1-\nu_3 + d/2 \ri }
{\Pi_{i} \G(\nu_i) } \no\\
\times 
\frac{ \G \le z_2 + z_3  + \nu_3 \ri  \G \le  \Sigma \nu_i - d/2 + z_3 + z_2 \ri }
{\G(d-\Sigma_i \nu_i)}, \\
D^{(z_2,z_3)}[1,1,1+\nu] \equiv D^{(z_2,z_3)}[1+\nu]
\end{eqnarray*}
In what follows we use the notation of Refs.\cite{Usyukina:1992jd,Usyukina:1993ch,Allendes:2012mr}, 
\begin{eqnarray*}
x \equiv \frac{p^2_1}{p^2_3},~~~ y \equiv \frac{p^2_2}{p^2_3},
\end{eqnarray*}
where the $d$-dimensional momenta $p_1,p_2,p_3$ satisfy the conservation law $p_1 + p_2 + p_3 = 0$ and 
are related to the $d$-dimensional momenta $q_1,q_2,q_3$ by the parametrization  
\begin{eqnarray}
p_1 = q_3 - q_2,~~~ p_2 = q_1 - q_3,~~~ p_3 = q_2 - q_1,  \label{p-q} 
\end{eqnarray}
which is chosen in a such way that $p_1$ appears to be a momentum that enters the one-loop triangle 
diagram in the vertex of the triangle which is opposite to the line with index $\nu_1$.

All the notation for MB integration are taken from Ref.\cite{Allendes:2012mr}, that is,  we absorb into the definition of the MB transform $D^{(z_2,z_3)}[\nu_1,\nu_2,\nu_3]$ 
of the three-point integral all the factors, except for a power of the square of the external momentum  $p^2_3.$ We do not write the powers of $i,$ 
assuming the Wick rotation is done. The contour of integration $C$ passes 
a bit on the left of the imaginary axis, separates left and right poles, and should be closed to the left infinity or to the right infinity. We choose to close the  
contour of integration in the complex plane to the right infinity. We omit the factor $1/(2\pi i)$ that accompanies each integration over an MB transform parameter. 
The inverse factor is generated in front of the residues.

In Ref. \cite{Allendes:2012mr} the procedure for work with the triangle-ladder diagrams in momentum space is described. Following this procedure, 
we use the notation 
\begin{eqnarray} \label{integral}
\int_n(\ve_1,\ve_2,\ve_3)
\end{eqnarray} 
for the momentum integral in $d=4-2\ve$ dimensions of the type which is depicted on l.h.s. of Fig. (\ref{figure10})
with the integral measure defined in Eq.~(\ref{k-measure}), but with  $n$ loops. All the formulae of this paper for the triangle ladder diagrams  coincide 
with the formulae of Ref. \cite{Allendes:2012mr} in the limit of the removing dimensional regularization, $\ve \rar 0.$
This means we change the sign of $\ve_1,\ve_2,\ve_3$ after the Fourier transformation on both the sides in  Fig. (\ref{figure10}).
The incoming momenta $p_1,p_2,p_3$ enter the triangle ladder 
diagram of the type depicted in Fig. (\ref{figure10}) in such a way that $p_3$ enters in the leftmost propagator, $p_1$ enters in the upper rightmost propagator, 
$p_2$ enters in the lower rightmost propagator. The external legs are amputated.
The $\ve_1,\ve_2,\ve_3$ terms appear in the indices of the lines for the first nine propagators on the l.h.s. of the diagram 
in momentum space as well as in position spaces. For example, the Fourier transform of the diagram on the l.h.s. of Fig. (\ref{figure10}) has the index $-\ve_3$ on the leftmost propagator,  
the indices  $1-\ve_1-\ve ,1-\ve_2,1-\ve_1$ on the upper side of the diagram, $1-\ve_2-\ve,1-\ve_1,1-\ve_2$ on the lower side of the diagram, and $1-\ve_3,1-\ve_3-\ve$ on the first 
two rungs. The residual lines have indices equal to $1-\ve$ on the rungs and 1 on the upper side and 1 on the lower side. These are the indices in 
momentum space.  The integral obtained at the end of these transformations is integral (\ref{integral}).

We apply the Fourier transformation to both the parts of Fig.~(\ref{figure04}), and in accordance with the rules described in the previous paragraphs of this section, we obtain the relations 
\begin{eqnarray*}
\int_2(\ve_1,\ve_2,\ve_3) = \frac{J}{(p_3^2)^{1-\ve_3 -\ve}}
\left[ \frac{1}{\ve_2\ve_3} \frac{1}{(p_2^2)^{-\ve_2}}  \frac{\G(1+\ve_1 - \ve)}{\G(1-\ve_1)} \frac{\G(1-\ve_1-\ve )}{\G(1+\ve_1)}  J(1,1,1-\ve_1-\ve)  + \right. \no\\  
\left.  \frac{1}{\ve_1\ve_2}  \frac{\G(1+\ve_3 - \ve)}{\G(1-\ve_3)} \frac{\G(1-\ve_3-\ve )}{\G(1+\ve_3)}  J(1,1,1+\ve_3-\ve) + \right.\\
\left. \frac{1}{\ve_1\ve_3} \frac{1}{(p_1^2)^{-\ve_1}} \frac{\G(1+\ve_2 - \ve)}{\G(1-\ve_2)} \frac{\G(1-\ve_2-\ve )}{\G(1+\ve_2)} J(1,1,1-\ve_2-\ve)
\right].            
\end{eqnarray*}
This expression  may be re-presented as  
\begin{eqnarray*}
\int_2(\ve_1,\ve_2,\ve_3) = \\
\frac{J}{(p_3^2)^{2-\ve}}
\left[ \frac{1}{\ve_2\ve_3} 
\frac{\G(1+\ve_1 - \ve)}{\G(1-\ve_1)} \frac{\G(1-\ve_1-\ve )}{\G(1+\ve_1)} \le\frac{ p_2^2}{p_3^2}\ri^{\ve_2}\oint_C du~dv~x^u~y^v   D^{(u,v)}[1-\ve_1-\ve] + \right.\no\\ 
\left. \frac{1}{\ve_1\ve_2} \frac{\G(1+\ve_3 - \ve)}{\G(1-\ve_3)} \frac{\G(1-\ve_3-\ve )}{\G(1+\ve_3)}\oint_C du~dv~x^u~y^v  D^{(u,v)}[1+\ve_3-\ve] + \right.\\
\left. \frac{1}{\ve_1\ve_3}   
\frac{\G(1+\ve_2 - \ve)}{\G(1-\ve_2)} \frac{\G(1-\ve_2-\ve )}{\G(1+\ve_2)} \le\frac{ p_1^2}{p_3^2}\ri^{\ve_1}  \oint_C du~dv~x^u~y^v  D^{(u,v)}[1-\ve_2-\ve] \right],   
\end{eqnarray*}
where $D^{(u,v)}[1-\ve_1]$ is a MB transform of a scalar ladder triangle diagram in the non-integer dimension $d=4-2\ve.$  We define the MB representation $M_2$ of  $\int_2(\ve_1,\ve_2,\ve_3)$ as 
\begin{eqnarray*}
\int_2(\ve_1,\ve_2,\ve_3) \equiv  \frac{1}{(p_3^2)^{2-\ve}} \oint_C du~dv~x^u~y^v   M_2^{(u,v)}[\ve_1,\ve_2,\ve_3].  
\end{eqnarray*}

For the three-rung triangle ladder, the diagrammatic relation of Fig.(10) in momentum space is  
\begin{eqnarray*}
\int_3(\ve_1,\ve_2,\ve_3) = \frac{ \G(1+\ve_1 - \ve)}{\G(1-\ve_1)} \frac{\G(1-\ve_1-\ve )}{\G(1+\ve_1)} \frac{1}{(p_1^2)^{\ve_1} (p_3^2)^{1+\ve_2}}
\frac{J}{\ve_2\ve_3} \int_2(\ve_1) + \\
 \frac{1}{\G(1-\ve)}\frac{1}{(p_1^2)^{\ve_1} (p_2^2)^{\ve_2} p_3^2} \frac{1}{\ve_1\ve_2}\int_2(\ve_1,\ve_2,\ve_3)  + \\
\frac{ \G(1+\ve_2 - \ve)}{\G(1-\ve_2)} \frac{\G(1-\ve_2-\ve )}{\G(1+\ve_2)} \frac{1}{(p_2^2)^{\ve_2} (p_3^2)^{1+\ve_1} }  \frac{J}{\ve_1\ve_3} \int_2(\ve_2).
\end{eqnarray*}
Without shifting variables of the contour integrals, we obtain 
\begin{eqnarray} \label{M3}
\int_3(\ve_1,\ve_2,\ve_3) = \no\\
\frac{J}{(p_3^2)^{3-\ve_3-\ve}} 
\left[ \frac{ \G(1+\ve_1 - \ve)}{\G(1-\ve_1)} \frac{\G(1-\ve_1-\ve )}{\G(1+\ve_1)} \frac{1}{\ve_2\ve_3} \le\frac{p_3^2}{p_1^2}\ri^{\ve_1}  
\oint_C du~dv~x^u~y^v M_2^{(u,v)}(\ve_1) + \right. \no\\
\left. 
J^{-1}\frac{1}{\G(1-\ve)}\frac{1}{\ve_1\ve_2} \le\frac{p_3^2}{p_1^2}\ri^{\ve_1} \le\frac{p_3^2}{p_2^2}\ri^{\ve_2}
\oint_C du~dv~x^u~y^v M_2^{(u,v)}(\ve_1,\ve_2,\ve_3) \right. \no\\
\left. + \frac{ \G(1+\ve_2 - \ve)}{\G(1-\ve_2)} \frac{\G(1-\ve_2-\ve )}{\G(1+\ve_2)} \frac{1}{\ve_1\ve_3} \le\frac{p_3^2}{p_2^2}\ri^{\ve_2} 
\oint_C du~dv~x^u~y^v M_2^{(u,v)}(\ve_2)\right].
\end{eqnarray}

The definition of the MB representation $M_3$ of  $\int_3(\ve_1,\ve_2,\ve_3)$ is
\begin{eqnarray*}
\int_3(\ve_1,\ve_2,\ve_3) \equiv  \frac{1}{(p_3^2)^{3-\ve_3-\ve}} \oint_C du~dv~x^u~y^v   M_3^{(u,v)}[\ve_1,\ve_2,\ve_3].  
\end{eqnarray*}

To obtain $\int_4(\ve_1,\ve_2,\ve_3)$ we have to integrate the l.h.s and each diagram on the r.h.s. of Fig. (\ref{figure10})  with three new propagators with indices $1-\ve,$ 1, $1-\ve$
in the p.s.r. and apply the Fourier transformation
\begin{eqnarray*}
\int_4(\ve_1,\ve_2,\ve_3) = \frac{ \G(1+\ve_1 - \ve)}{\G(1-\ve_1)} \frac{\G(1-\ve_1-\ve )}{\G(1+\ve_1)} \frac{1}{(p_3^2)^{1+\ve_2}}\frac{J}{\ve_2\ve_3} \int_3(\ve_1) + \\
\frac{1}{p_3^2} \frac{1}{\G(1-\ve)}\frac{1}{\ve_1\ve_3}\int_3(\ve_1,\ve_2,\ve_3)  + 
\frac{ \G(1+\ve_2 - \ve)}{\G(1-\ve_2)} \frac{\G(1-\ve_2-\ve )}{\G(1+\ve_2)} \frac{1}{(p_3^2)^{1+\ve_1} }  \frac{J}{\ve_1\ve_3} \int_3(\ve_2) .
\end{eqnarray*}
This formula may be re-written as
\begin{eqnarray*}
\int_4(\ve_1,\ve_2,\ve_3) = \\ 
\frac{ \G(1+\ve_1 - \ve)}{\G(1-\ve_1)} \frac{\G(1-\ve_1-\ve )}{\G(1+\ve_1)}  \frac{J}{\ve_2\ve_3} \frac{1}{ (p_3^2)^{1+\ve_2} (p_3^2)^{3+\ve_1 -\ve} } \oint_C du~dv~x^u~y^v M_3^{(u,v)}(\ve_1) + \\
\frac{1}{\G(1-\ve)}\frac{1}{\ve_1\ve_2} \frac{1}{(p_3^2)^{4-\ve_3 -\ve}} \oint_C du~dv~x^u~y^v M_3^{(u,v)}(\ve_1,\ve_2,\ve_3) + \\
\frac{ \G(1+\ve_2 - \ve)}{\G(1-\ve_2)} \frac{\G(1-\ve_2-\ve )}{\G(1+\ve_2)}  \frac{J}{\ve_1\ve_3} \frac{1}{(p_3^2)^{1+\ve_1} (p_3^2)^{3+\ve_2 - \ve}} \oint_C du~dv~x^u~y^v M_3^{(u,v)}(\ve_2) = \\
\frac{J}{ (p_3^2)^{4-\ve_3-\ve}}
\left[\frac{ \G(1+\ve_1 - \ve)}{\G(1-\ve_1)} \frac{\G(1-\ve_1-\ve )}{\G(1+\ve_1)}  \frac{1}{\ve_2\ve_3} \oint_C du~dv~x^u~y^v M_3^{(u,v)}(\ve_1) 
+ \right. \\
\left. J^{-1}\frac{1}{\G(1-\ve)}\frac{1}{\ve_1\ve_2} 
\oint_C du~dv~x^u~y^v M_3^{(u,v)}(\ve_1,\ve_2,\ve_3) + \right.\\
\left. \frac{ \G(1+\ve_2 - \ve)}{\G(1-\ve_2)} \frac{\G(1-\ve_2-\ve )}{\G(1+\ve_2)}  \frac{1}{\ve_1\ve_3}  \oint_C du~dv~x^u~y^v 
M_3^{(u,v)}(\ve_2)\right].
\end{eqnarray*}

To calculate the limit in which $\ve_1,\ve_2$ are vanishing, we need to replace the function $D[1,1,1-\ve-\delta]$ with a linear combination of  other two MB transforms
\begin{eqnarray*}
\oint_C du~dv~x^u~y^v D^{(u,v)}[1,1,1-\delta - \ve] = \no\\
\frac{1}{2} \oint_C du~dv~x^u~y^v\Bigl(x^{\delta}D^{(u,v)}[1-\delta,1+\delta,1-\ve] + y^{\delta}D^{(u,v)}[1+\delta,1-\delta,1-\ve] \Bigr) \times \\
\times\frac{\G(1-\delta)\G(1+\delta)\G^2(1-\ve)}{\G(1-\delta-\ve)\G(1+\delta-\ve)}.
\end{eqnarray*}
Such a trick allows us to find the limit of vanishing $\ve_1,\ve_2$ for $M_n.$ For example, for $M_2$ we have  
\begin{eqnarray*}
\oint_C du~dv~x^u~y^v M_2^{(u,v)}(\ve_1,\ve_2,\ve_3) = \no\\
J\oint_C du~dv~x^u~y^v \Bigl[ \frac{ \G(1+\ve_1 - \ve)}{\G(1-\ve_1)} \frac{\G(1-\ve_1-\ve )}{\G(1+\ve_1)} 
\frac{1}{\ve_2\ve_3} y^{\ve_2} D^{(u,v)}[1-\ve_1-\ve] + \Bigr.\\
\Bigl. \frac{ \G(1+\ve_3 - \ve)}{\G(1-\ve_3)} \frac{\G(1-\ve_3-\ve )}{\G(1+\ve_3)} \frac{1}{\ve_1\ve_2}  D^{(u,v)}[1+\ve_3 -\ve] + \Bigr.\\
\Bigl. \frac{ \G(1+\ve_2 - \ve)}{\G(1-\ve_2)} \frac{\G(1-\ve_2-\ve )}{\G(1+\ve_2)}  \frac{1}{\ve_1\ve_3} x^{\ve_1} D^{(u,v)}[1-\ve_2-\ve] \Bigr] = \no\\
\frac{J}{2}\oint_C du~dv~x^u~y^v \left[ \frac{1}{\ve_2\ve_3} y^{\ve_2} \times \right.\\ 
\left.\times  \frac{x^{\ve_1}\Gamma(-u-\ve_1)\Gamma(-v+\ve_1) + y^{\ve_1}\Gamma(-u+\ve_1)\Gamma(-v-\ve_1)}{\G(1-\ve_1 - \ve)\G(1+\ve_1-\ve)}  
\frac{ \G(1+\ve_1 - \ve)}{\G(1-\ve_1)}  \frac{\G(1-\ve_1-\ve )}{\G(1+\ve_1)}  \right. \no\\
\left. 
+ \frac{1}{\ve_1\ve_2} \frac{x^{-\ve_3}\Gamma(-u+\ve_3)\Gamma(-v-\ve_3) + y^{-\ve_3}\Gamma(-u-\ve_3)\Gamma(-v+\ve_3)} {\G(1-\ve_3 - \ve)\G(1+\ve_3-\ve)}  
\frac{ \G(1+\ve_3 - \ve)}{\G(1-\ve_3)}  \frac{\G(1-\ve_3-\ve )}{\G(1+\ve_3)} \right. \no\\
\left. 
+ \frac{1}{\ve_1\ve_3} x^{\ve_1} \frac{x^{\ve_2}\Gamma(-u-\ve_2)\Gamma(-v+\ve_2) + y^{\ve_2}\Gamma(-u+\ve_2)\Gamma(-v-\ve_2)}{\G(1-\ve_2 - \ve)\G(1+\ve_2-\ve)} \times\right.\\ 
\left. \times\frac{ \G(1+\ve_2 - \ve)}{\G(1-\ve_2)} \frac{\G(1-\ve_2-\ve )}{\G(1+\ve_2)}\right]\G(-u)\G(-v) \G(1+u+v)\G(1+u+v-\ve) = \\
\frac{J}{2}\oint_C du~dv~x^u~y^v \left[ \frac{a(\ve_1)}{\ve_2\ve_3} y^{\ve_2} \times \right.\\ 
\left.\times  \left[x^{\ve_1}\Gamma(-u-\ve_1)\Gamma(-v+\ve_1) + y^{\ve_1}\Gamma(-u+\ve_1)\Gamma(-v-\ve_1)\right] \right. \no\\
\left. 
+ \frac{a(\ve_3)}{\ve_1\ve_2} \left[x^{-\ve_3}\Gamma(-u+\ve_3)\Gamma(-v-\ve_3) + y^{-\ve_3}\Gamma(-u-\ve_3)\Gamma(-v+\ve_3)\right]  \right. \no\\
\left. 
+ \frac{a(\ve_2)}{\ve_1\ve_3} x^{\ve_1} \left[x^{\ve_2}\Gamma(-u-\ve_2)\Gamma(-v+\ve_2) + y^{\ve_2}\Gamma(-u+\ve_2)\Gamma(-v-\ve_2)\right] \right]\times\\
\G(-u)\G(-v) \G(1+u+v)\G(1+u+v-\ve) 
\end{eqnarray*}
The definition of $a(\ve_i)$ remains the same as in Ref. \cite{Allendes:2012mr}, 
\begin{eqnarray*}
a(\delta) = \left[ \G(1-\delta)\G(1+\delta)\right]^{-1}. 
\end{eqnarray*}
Taking the double uniform limit $\ve_1 \rar 0, \ve_2 \rar 0,$ we obtain 
\begin{eqnarray} \label{M2-limit}
\lim_{\ve_2 \rar 0, \ve_1 \rar 0} \oint_C du~dv~x^u~y^v M_2^{(u,v)}(\ve_1,\ve_2,\ve_3) = \no\\
\frac{1}{\G^3(1-\ve)}\oint_C du~dv~x^u~y^v \G(-u)\G(-v) \G(1+u+v)\G(1+u+v-\ve)\times \no\\
\times\left[\frac{3}{2}\le a(\ve)\G(-u-\ve)\G(-v + \ve)\ri^{(2)}_0 + \right. \no\\
\left. \frac{3}{2}\ln\frac{x}{y}\le a(\ve)\G(-u-\ve)\G(-v + \ve)\ri'_0 + \frac{1}{4}\ln^2\frac{x}{y}\G(-u)\G(-v)\right].
\end{eqnarray}
From Eq. (\ref{M3}), if we collect all the terms on the r.h.s. in the same integrand, we obtain 
\begin{eqnarray*}
\oint_C du~dv~x^u~y^v M_3^{(u,v)}(\ve_1,\ve_2,\ve_3) =  \\
\oint_C du~dv~x^u~y^v \left[\frac{1}{\G(1-\ve)}\frac{1}{\ve_1\ve_2} x^{-\ve_1} y^{-\ve_2} M_2^{(u,v)}(\ve_1,\ve_2,\ve_3) + \right.\\ 
\left. + \frac{J}{\ve_2\ve_3}\frac{ \G(1+\ve_1 - \ve)}{\G(1-\ve_1)}  \frac{\G(1-\ve_1-\ve )}{\G(1+\ve_1)} x^{-\ve_1} M_2^{(u,v)}(\ve_1) \right.\\ 
\left. +  \frac{J}{\ve_1\ve_3} \frac{ \G(1+\ve_2 - \ve)}{\G(1-\ve_2)}  \frac{\G(1-\ve_2-\ve )}{\G(1+\ve_2)} y^{-\ve_2} M_2^{(u,v)}(\ve_2)\right]. 
\end{eqnarray*} 
Taking the double uniform limit of vanishing $\ve_1,\ve_2$ we obtain  
\begin{eqnarray} \label{M3-limit}
\lim_{\ve_2 \rar 0, \ve_1 \rar 0}  \oint_C du~dv~x^u~y^v M_3^{(u,v)}(\ve_1,\ve_2,\ve_3) = \no\\
\frac{1}{\G^4(1-\ve)}\oint_C du~dv~x^u~y^v \G(-u)\G(-v) \G(1+u+v)\G(1+u+v-\ve)\times\no\\
\times\left[\frac{5}{12} \le a(\ve)\G(-u-\ve)\G(-v + \ve)\ri^{(4)}_0 + \right. \no\\
\left.
\frac{5}{6}\ln \frac{x}{y} \le a(\ve)\G(-u-\ve)\G(-v + \ve)\ri^{(3)}_0 + 
\frac{1}{2}\ln^2 \frac{x}{y} \le a(\ve)\G(-u-\ve)\G(-v + \ve) \ri^{(2)}_0 \right. \no\\
\left.
+ \frac{1}{12}\ln^3 \frac{x}{y} \le a(\ve)\G(-u-\ve)\G(-v + \ve)\ri'_0  \right]. 
\end{eqnarray}

\section{Index on the first rung of the triangle ladder}

\begin{figure}[h!!!] 
\centering\includegraphics[scale=.5]{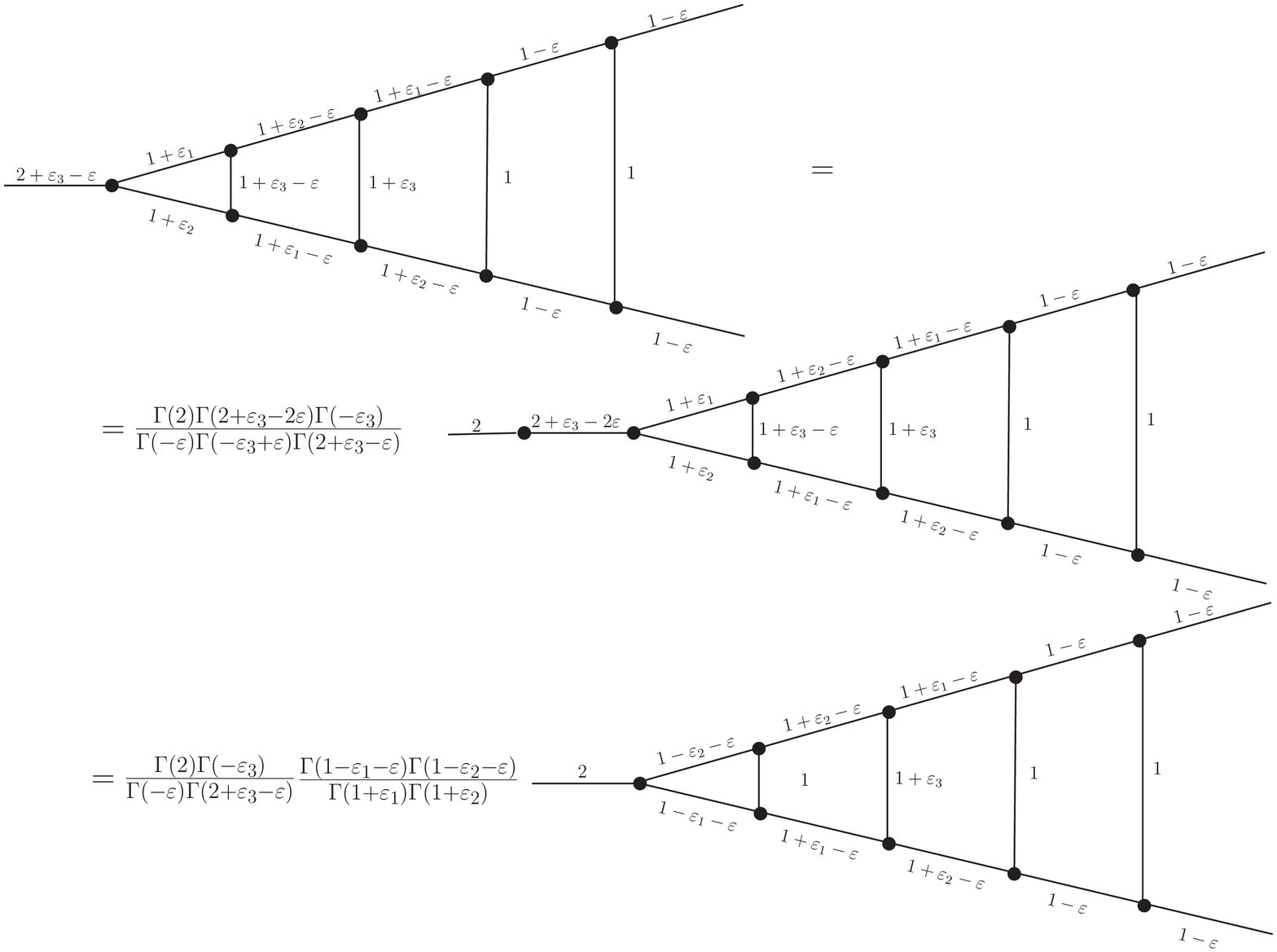}
\vspace{-0.4cm}
 \caption{\footnotesize  Change of the index on the first rung of the diagram in Fig. (\ref{figure11})}
\label{figure12}
 \end{figure}

\begin{figure}[h!!!] 
\centering\includegraphics[scale=.5]{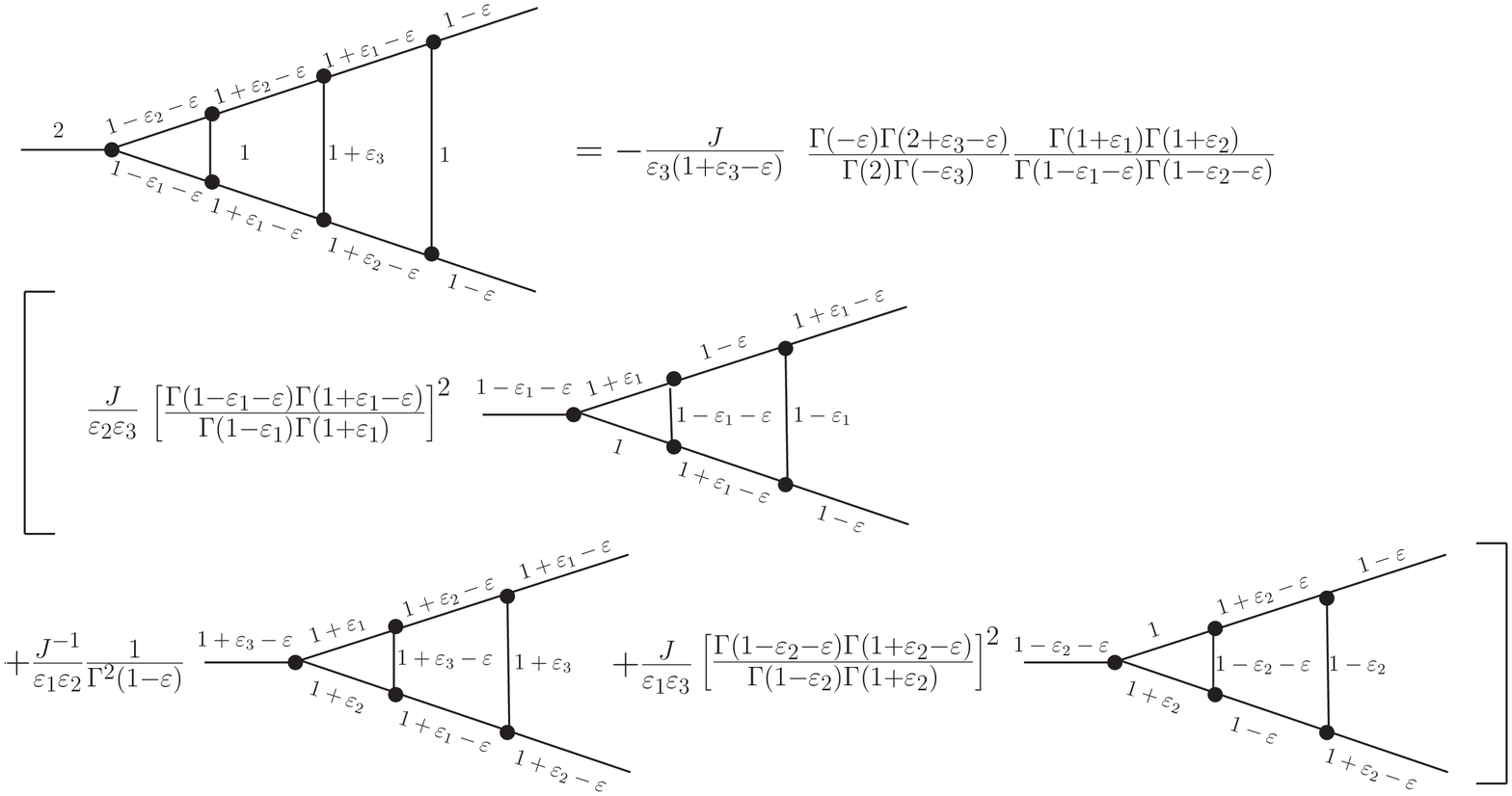}
\vspace{-0.4cm}
 \caption{\footnotesize  Loop reduction relation for the modified diagram of three rungs}
\label{figure13}
 \end{figure}

\begin{figure}[h!!!] 
\centering\includegraphics[scale=.5]{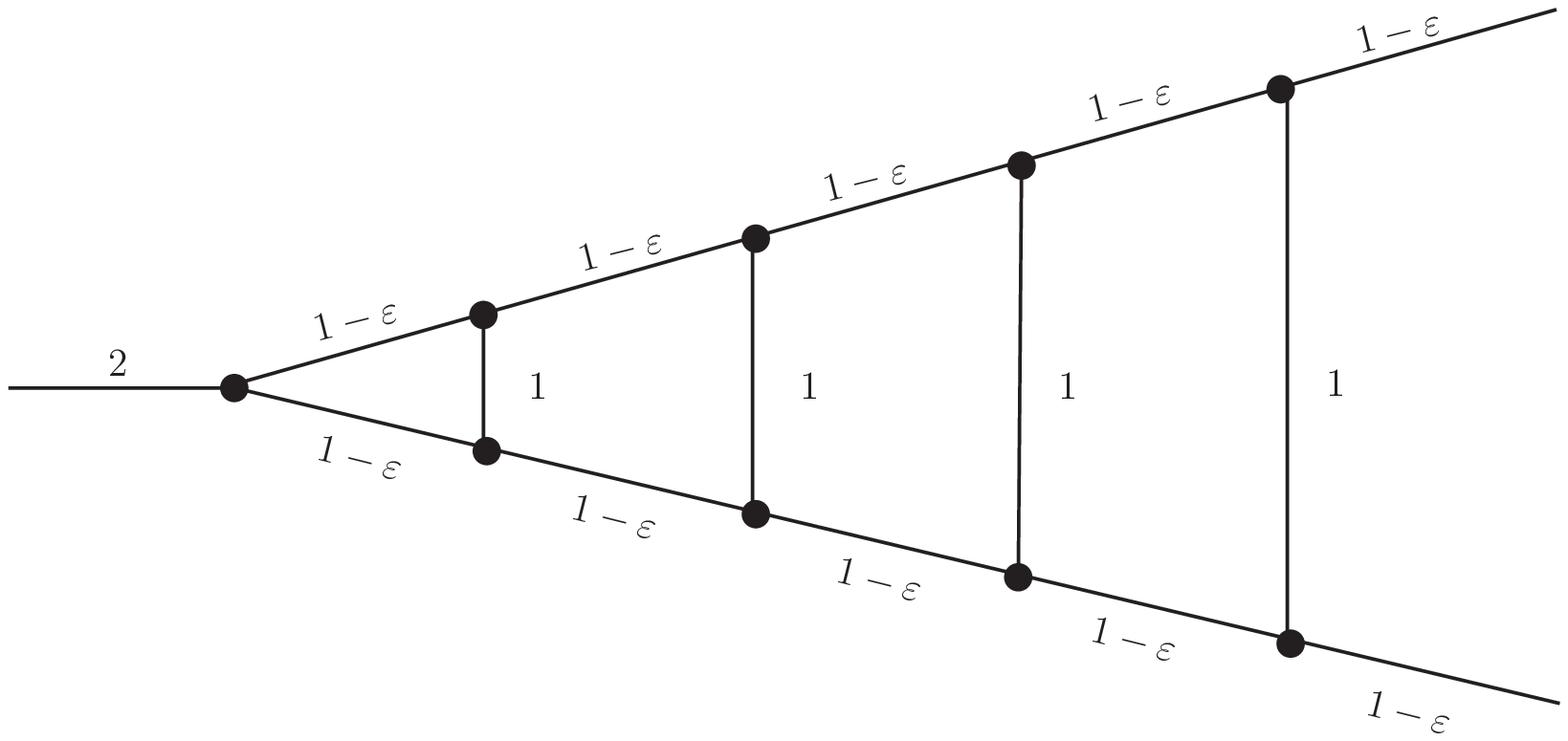}
\vspace{-0.4cm}
 \caption{\footnotesize  The indices in the position space after taking the limit of vanishing the analytic regularization parameters}
\label{figure14}
 \end{figure}

\begin{figure}[h!!!] 
\centering\includegraphics[scale=.5]{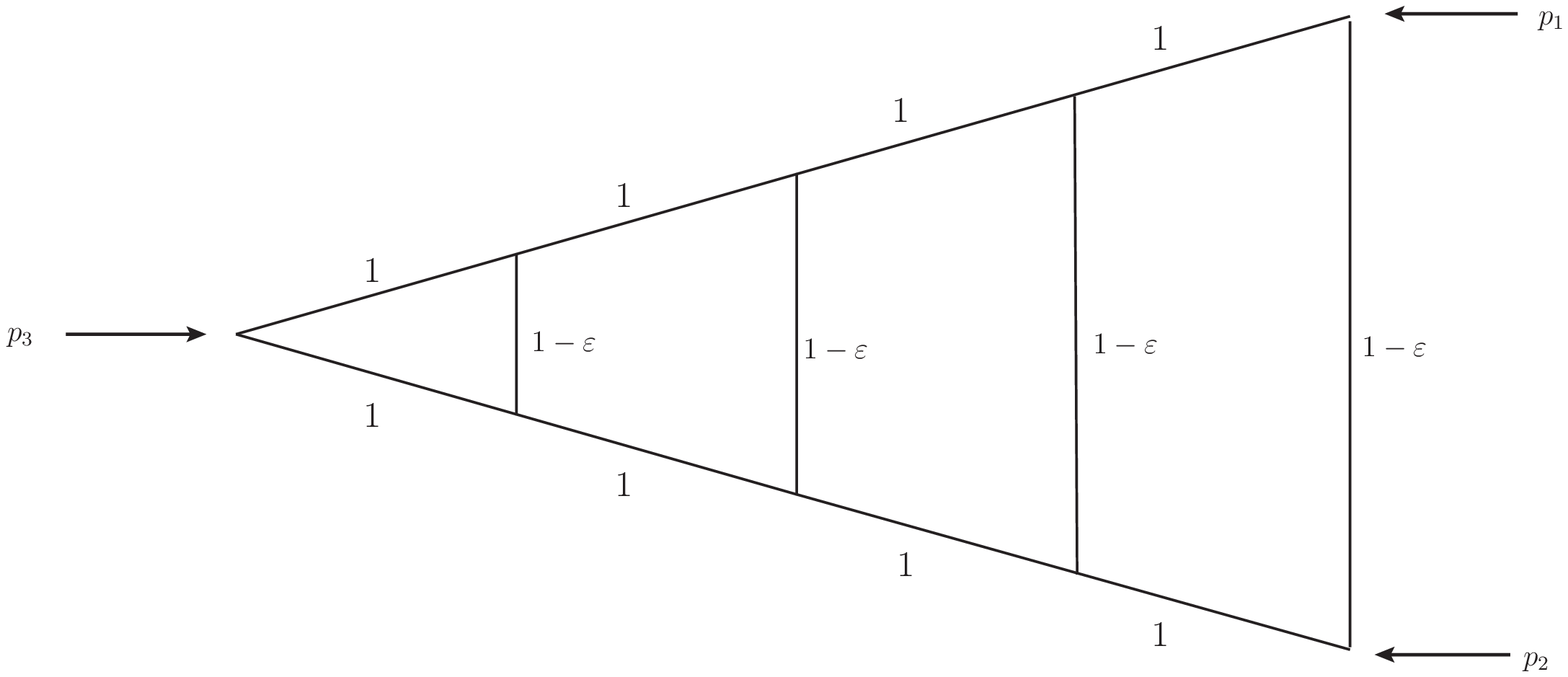}
\vspace{-0.4cm}
 \caption{\footnotesize  Indices of Fig. (\ref{figure14}) in the momentum space}
\label{figure15}
 \end{figure}

If we take a limit when the parameters of analytical regularization $\ve_1,\ve_2$ are vanishing, the family of  diagrams depicted in  Fig. (\ref{figure11})
will be a bit asymmetric, since on the first rung we have  the index $1-\ve$ while on the rest of rungs we have the indices 1. To treat this asymmetry we 
make the transformation of this family of diagrams depicted in  Fig. (\ref{figure11}) by using again the trick of inserting the point into a propagator.
This is done in order to get the uniqueness condition at the leftmost internal vertex of the diagram. The trick is explained in Fig. (\ref{figure12}). 
As the result, when we remove the analytical regularization, we gain the diagram in which all the rungs have  index 1 in  position space. 
The recursive relation for the modified diagram with three rungs is depicted in Fig. (\ref{figure13}). 
After the transformation explained in Fig. (\ref{figure12}), we have the tool to calculate 
the triangle ladder diagrams in the p.s.r. with the index distribution, depicted in Fig. (\ref{figure14}).  Then, after applying the Fourier transformation, 
the index $1-\ve$ transforms to 1 and vice verse.  Example of the four-rung diagram in the m. s. r. is given in  Fig. (\ref{figure15}).

\begin{figure}[h!!!] 
\centering\includegraphics[scale=.5]{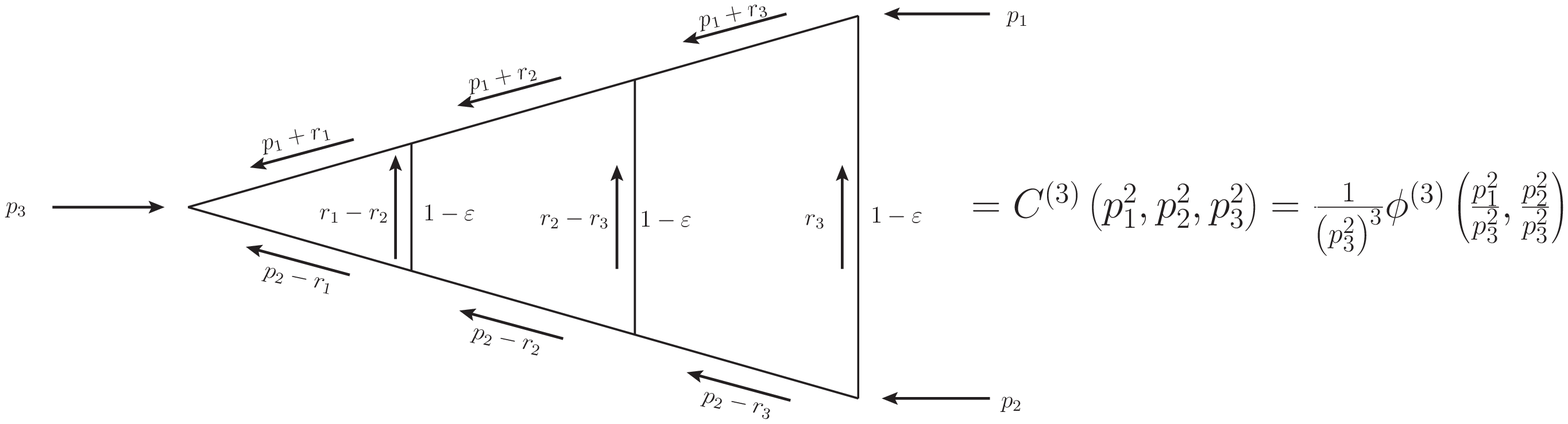}
\vspace{-0.4cm}
 \caption{\footnotesize  Momenta distribution in the three-rung ladder}
\label{figure16}
 \end{figure}

The result of calculation of the triangle ladder diagrams in the m.s.r. in $d=4$ is UD functions \cite{Usyukina:1992jd,Usyukina:1993ch,Broadhurst:2010ds}. 
Their properties, in particular their  invariance with respect to the Fourier transformation, have been studied in 
Refs. \cite{Kondrashuk:2008ec,Kondrashuk:2008xq,Kondrashuk:2009us, Allendes:2009bd} and their MB transforms 
have been studied in Refs. \cite{Allendes:2009bd,Allendes:2012mr}.    In  Fig. (\ref{figure16}) 
we introduce a parametrization $\phi^{(n)}(x,y)$ for the momentum integral of the triangle ladder diagram in consideration. It has been done in analogy with  the case of $d=4.$  
The dimensionless function $\phi^{(3)}(x,y)$ that appears on the r.h.s. of Fig. (\ref{figure16}) 
is not the UD function $\Phi^{(3)}(x,y)$ that appears for a similar diagram with all the indices in the m.s.r. equal to 1 in the case of integer dimension $d=4.$ 
In the limit of removing the dimensional regularization we may write the relation 
\begin{eqnarray*}
\lim_{\ve \rightarrow 0} \phi^{(3)}(x,y) = \Phi^{(3)}(x,y),
\end{eqnarray*}
however, this limit has a sense off shell only, since for the on-shell momenta this limit does not exist. All the triangle ladders are related by the integral relation
\begin{eqnarray*}
C^{(n)}(p_1^2,p_2^2,p_3^2) =  \int~d^{4-2\ve}r_n~\frac{C^{(n-1)}((p_1 + r_n)^2,(p_2 - r_n)^2, p_3^2 )}
{(p_1 + r_n)^2 (p_2 - r_n)^2 (r_n^2)^{1-\ve}}, 
\end{eqnarray*}
which is a generalization of the four-dimensional relations \cite{Usyukina:1992jd,Usyukina:1993ch,Kondrashuk:2009us} and is a consequence of the ladder-like 
structure of this family of diagrams.

To calculate $\phi^{(n)}(x,y)$ we need to relate the m.s.r. of the l.h.s. and r.h.s. of Fig. (\ref{figure12}). From Fig. (\ref{figure12})  we read off the equation 
\begin{eqnarray*}
\int_4(\ve_1,\ve_2,\ve_3) = \frac{ \G(1+\ve_1)\G(1+\ve_2)\G(1+\ve_3-\ve)\G(1-\ve)}{\G(1-\ve_1-\ve)\G(1-\ve_2-\ve)\G(1-\ve_3)} \frac{1}{(p_3^2)^{-\ve_3 -\ve}}\int'_4(\ve_1,\ve_2,\ve_3),
\end{eqnarray*}
where integral $\int'_4(\ve_1,\ve_2,\ve_3)$ corresponds to the r.h.s. of Fig. (\ref{figure12}) after making Fourier transformation and changing signs of the 
parameters of analytical regularization $\ve_1,\ve_2,\ve_3.$ The external legs remain amputated.  In the limit when parameters  $\ve_1,\ve_2$  are vanishing, we obtain 
\begin{eqnarray*}
\lim_{\ve_2 \rar 0, \ve_1 \rar 0} \int_4(\ve_1,\ve_2,\ve_3) =  \frac{1}{(p_3^2)^{-\ve}} \lim_{\ve_2 \rar 0, \ve_1 \rar 0}\int'_4(\ve_1,\ve_2,\ve_3).
\end{eqnarray*}
From this equation and from definition in Fig. (\ref{figure16}) we conclude
\begin{eqnarray*}
C^{(4)}(p_1^2,p_2^2,p_3^2) =  \lim_{\ve_2 \rar 0, \ve_1 \rar 0}\int'_4(\ve_1,\ve_2,\ve_3) =   (p_3^2)^{-\ve}\lim_{\ve_2 \rar 0, \ve_1 \rar 0} \int_4(\ve_1,\ve_2,\ve_3).
\end{eqnarray*}
This relation may be generalized to an arbitrary $n$ as 
\begin{eqnarray} \label{Cn}
C^{(n)}(p_1^2,p_2^2,p_3^2) =  \lim_{\ve_2 \rar 0, \ve_1 \rar 0}\int'_n(\ve_1,\ve_2,\ve_3) =   (p_3^2)^{-\ve}\lim_{\ve_2 \rar 0, \ve_1 \rar 0} \int_n(\ve_1,\ve_2,\ve_3).
\end{eqnarray}
From Eqs.(\ref{M2-limit}),(\ref{M3-limit}) and from Eq.(\ref{Cn}) we obtain  
\begin{eqnarray*}
\phi^{(2)}(x,y) =  \frac{1}{\G^3(1-\ve)}\oint_C du~dv~x^u~y^v \G(-u)\G(-v) \G(1+u+v)\G(1+u+v-\ve)\times\\
\times \left[\frac{3}{2}\le a(\ve)\G(-u-\ve)\G(-v + \ve)\ri^{(2)}_0 + 
\frac{3}{2}\ln\frac{x}{y}\le a(\ve)\G(-u-\ve)\G(-v + \ve)\ri'_0 \right.\\
\left. + \frac{1}{4}\ln^2\frac{x}{y}\G(-u)\G(-v)\right],
\end{eqnarray*}
and 
\begin{eqnarray*}
\phi^{(3)}(x,y) =
\frac{1}{\G^4(1-\ve)}\oint_C du~dv~x^u~y^v \G(-u)\G(-v) \G(1+u+v)\G(1+u+v-\ve) \times\\
\times \left[\frac{5}{12} \le a(\ve)\G(-u-\ve)\G(-v + \ve)\ri^{(4)}_0 + 
\frac{5}{6}\ln \frac{x}{y} \le a(\ve)\G(-u-\ve)\G(-v + \ve)\ri^{(3)}_0 \right. \no\\
\left. + \frac{1}{2}\ln^2 \frac{x}{y} \le a(\ve)\G(-u-\ve)\G(-v + \ve) \ri^{(2)}_0  + \frac{1}{12}\ln^3 \frac{x}{y} \le a(\ve)\G(-u-\ve)\G(-v + \ve)\ri'_0  \right]. 
\end{eqnarray*}
However, our purpose is the box ladders.  In what follows, we will show how the triangle ladder of  Fig. (\ref{figure15}) transforms to the corresponding 
box ladder with the same index distribution  in the m.s.r., namely, to the ladders with indices  $1-\ve$  on the rungs and with index 1 in the rest of lines.

\section{Transformation of the triangle ladders to the box ladders}

To show how the triangle ladders of Fig. (\ref{figure15}) can be transformed to the box ladders we follow the methods described in Ref. \cite{Kondrashuk:2009us} 
for the case of $d=4.$

\begin{figure}[h!!] 
\centering\includegraphics[scale=.5]{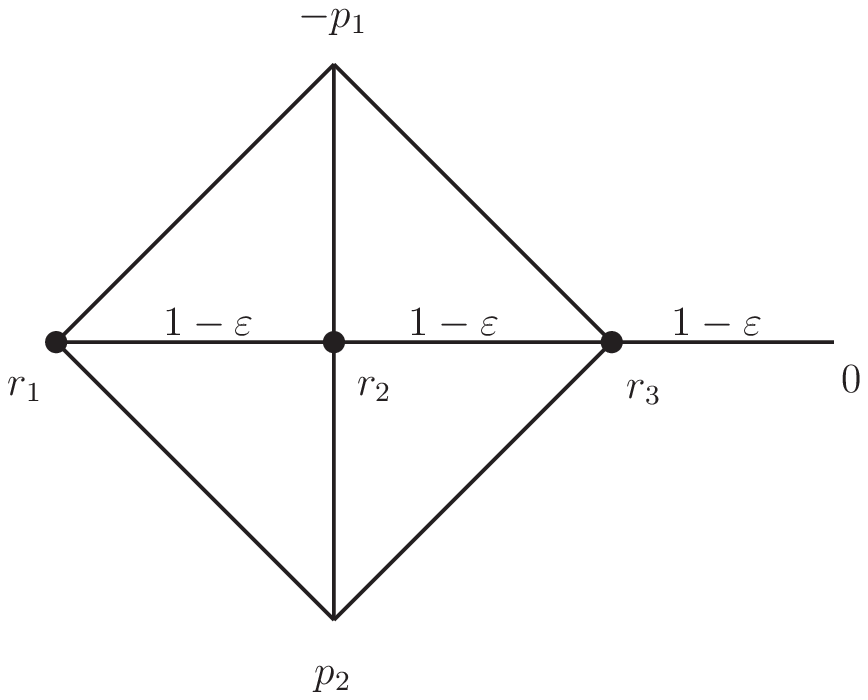}
\vspace{-0.4cm}
 \caption{\footnotesize  Dual space image of the momentum diagram in  Fig. (\ref{figure16})}
\label{figure17}
 \end{figure}

\begin{figure}[h!!!] 
\centering\includegraphics[scale=.5]{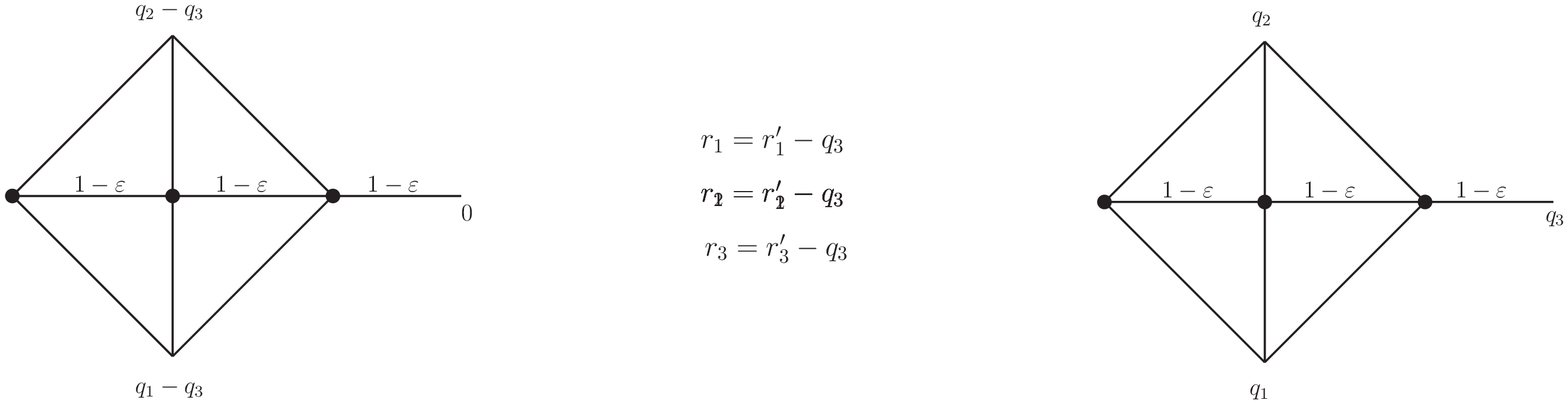}
\vspace{-0.4cm}
 \caption{\footnotesize  Shift of the integration variables}
\label{figure18}
 \end{figure}

\begin{figure}[h!!] 
\centering\includegraphics[scale=.5]{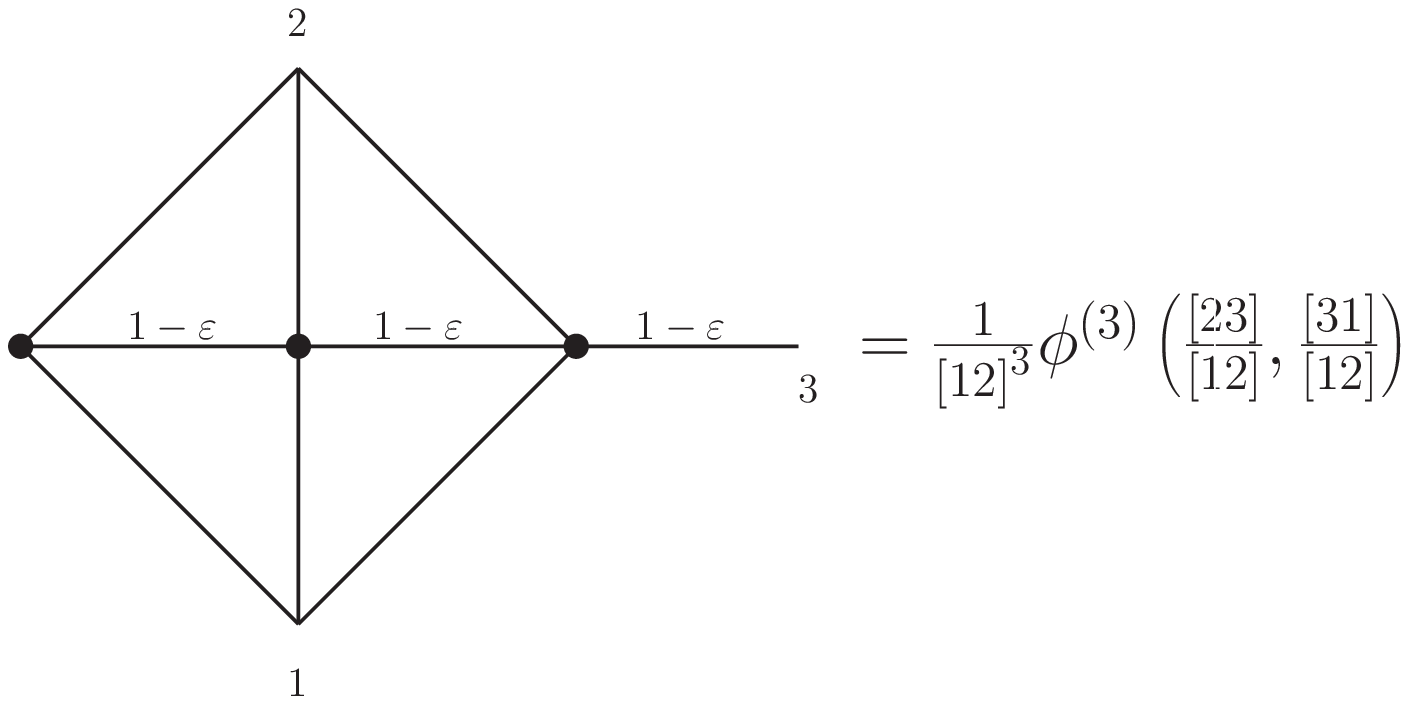}
\vspace{-0.4cm}
 \caption{\footnotesize  Parametrization for the three-loop momenta diagram of Fig. (\ref{figure16}) }
\label{figure19}
 \end{figure}

\begin{figure}[h!!!] 
\centering\includegraphics[scale=.5]{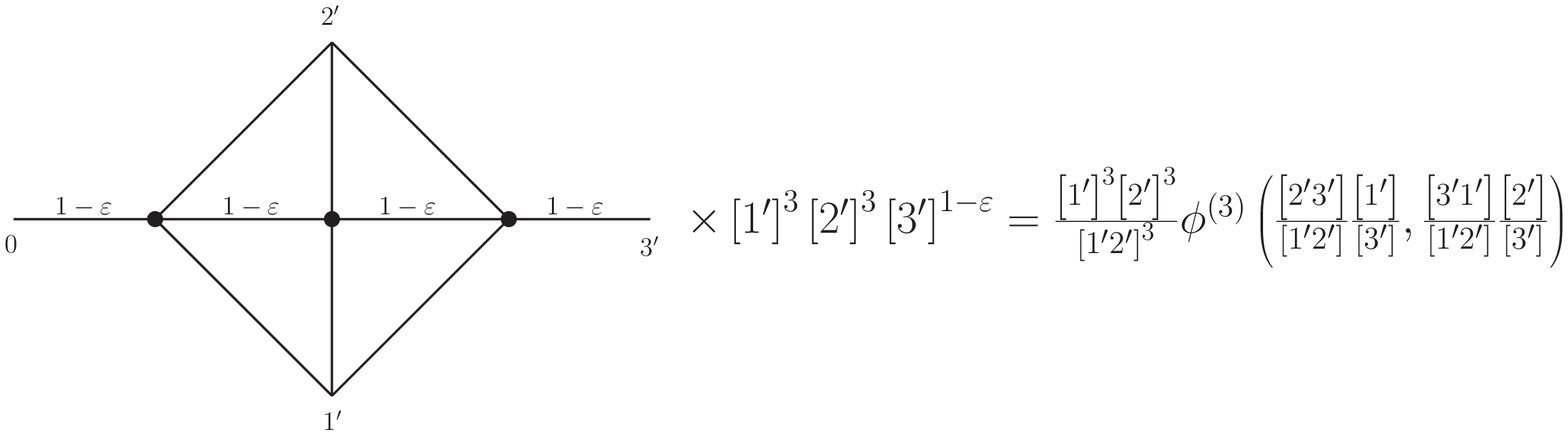}
\vspace{-0.4cm}
 \caption{\footnotesize  Conformal transformation of Fig. (\ref{figure19}) in the dual space}
\label{figure20}
 \end{figure}

\begin{figure}[h!!!] 
\centering\includegraphics[scale=.5]{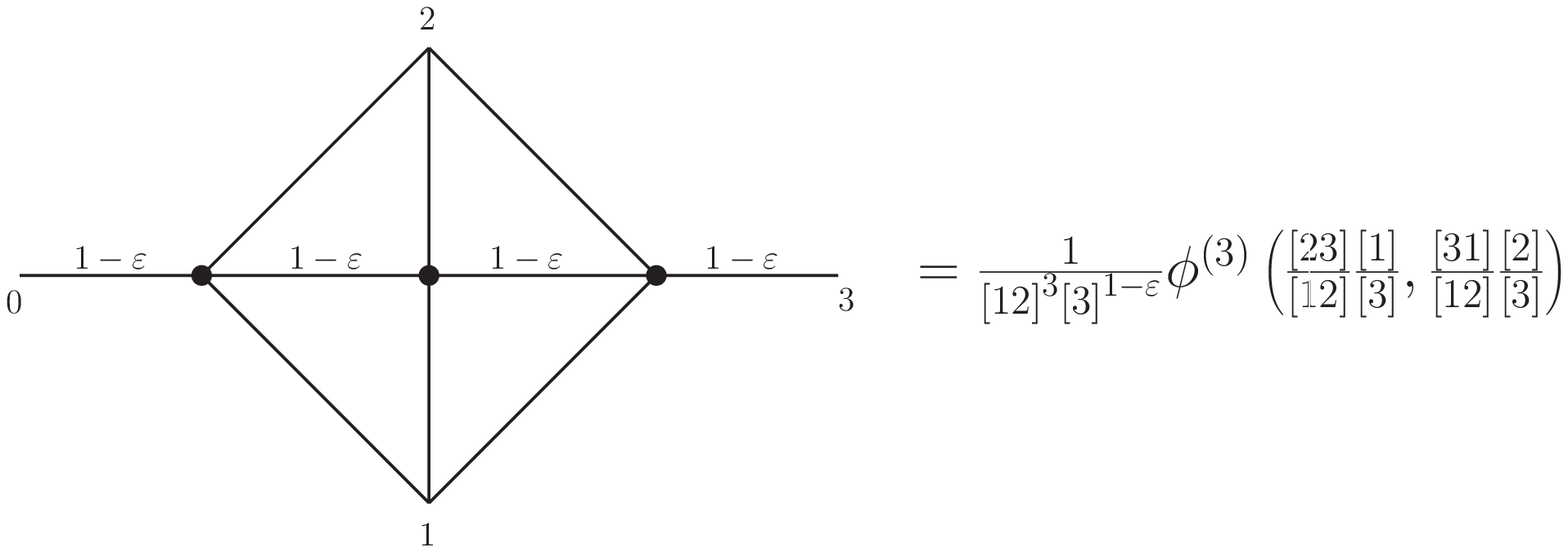}
\vspace{-0.4cm}
 \caption{\footnotesize  Result for the diamond diagram in the dual space }
\label{figure21}
 \end{figure}

\begin{figure}[h!!!] 
\centering\includegraphics[scale=.5]{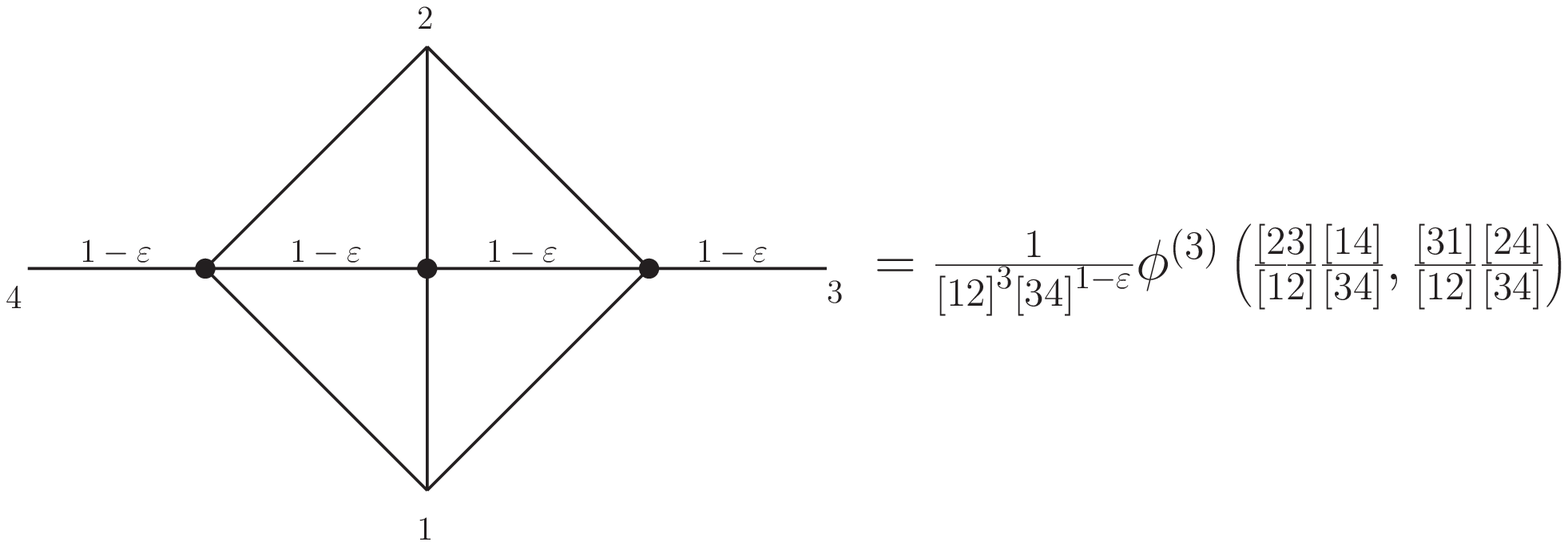}
\vspace{-0.4cm}
 \caption{\footnotesize  Shift of the variables in the dual diagram }
\label{figure22}
 \end{figure}

\begin{figure}[h!!!] 
\centering\includegraphics[scale=.5]{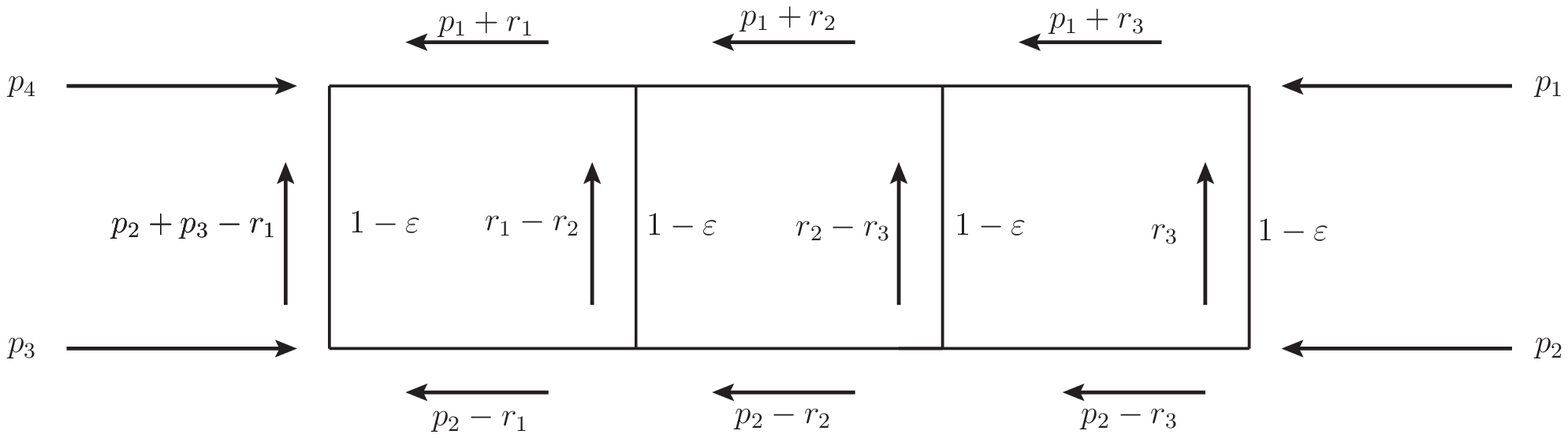}
\vspace{-0.4cm}
 \caption{\footnotesize  Momentum space image of the dual space diamond }
\label{figure23}
 \end{figure}

\begin{figure}[h!!!] 
\centering\includegraphics[scale=.5]{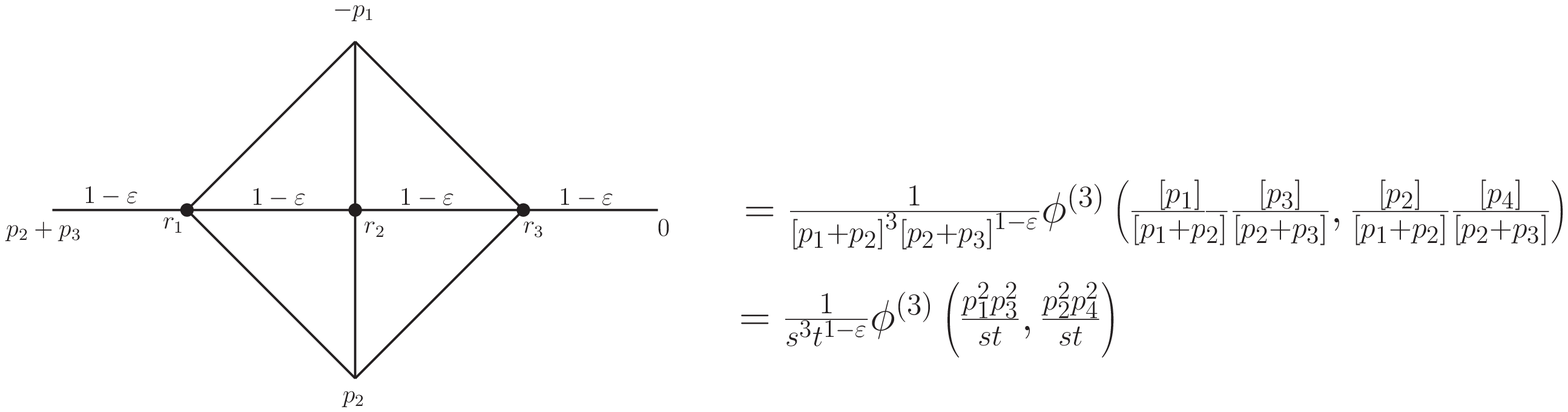}
\vspace{-0.4cm}
 \caption{\footnotesize  Result for the three-loop box ladder diagram of  (\ref{figure23}) from the dual space diamond of Fig. (\ref{figure22}) }
\label{figure24}
 \end{figure}

We go now to the dual space representation (the d.s.r.) of the momentum integrals which in case of the example in Fig. (\ref{figure16}) has a form depicted in  Fig. (\ref{figure17}).
The review of this construction has been done in Ref. \cite{Kondrashuk:2009us}. Till the end of this paper we follow the lines of Ref. \cite{Kondrashuk:2009us}.
In  Fig. (\ref{figure17}) only the indices distinct from 1 are written. If the index is 1, we omit it in all the next figures.  
The lines without any index mean that these are lines with index 1 in momentum space. Then, we do a redefinition (\ref{p-q}).
The diagram of Fig. (\ref{figure17}) takes the form depicted on the l.h.s. of Fig. (\ref{figure18}). Then we do a shift of variables of integration in the internal vertices of the dual graph.
This shift of the variables is written in the center of Fig. (\ref{figure18}). 
As the result, we obtain the figure on the r.h.s. of Fig. (\ref{figure18}). 
The new variables of integrations (after the shift) $r'_1,r'_2,r'_3$ are related to the initial variables
as  $r_1 = r'_1 - q_3,$ $r_2 = r'_2 - q_3,$ and $r_3 = r'_3 - q_3.$  The l.h.s. of Fig. (\ref{figure18}) and  the r.h.s. of Fig. (\ref{figure18})
are equal, and they are equal to the l.h.s. of Fig. (\ref{figure19})  in which $N=1,2,3$ stands for $q_N=q_1,q_2,q_3.$

The diagram in Fig.(\ref{figure19}) in dual space can be viewed as a diagram in position space generated for another field theory
defined in the same non-integer number of $d=4-2\ve$ dimensions.  The letters $q_1,q_2,q_3$ corresponding to the external points of the d.s.r. can be replaced with 
the letters $x_1,x_2,x_3$ corresponding to any three distinct points in position space of this auxiliary scalar field theory. 
This correspondence allows the interpretation of the dual space diagram as a position space diagram with massless scalar propagators.
The concise notation of  Ref.\cite{Cvetic:2006iu} for the space-time intervals is used, namely   
$[Nx]= (x_N - x)^2$  and $[12]= (x_1 - x_2)^2 ,$ that is, $N=1,2,3$ stands for $x_N=x_1,x_2,x_3,$ in complete analogy to the case of $d=4$ of Ref. \cite{Kondrashuk:2009us}.
In this notation and in according to the definition of the three-rung triangle ladder function of Fig.(\ref{figure16}) the result for the diagram of 
Fig. (\ref{figure19}) should be written as 
\begin{eqnarray*}
\frac{1}{[12]^3} \phi^{(3)}\le \frac{[23]}{[12]},\frac{[31]}{[12]}\ri.
\end{eqnarray*}
This identity is depicted in  Fig.(\ref{figure19}).

According to the interpretation of the previous paragraph, we treat the internal vertices of integration in the dual space graph in Fig.(\ref{figure19})
as new variables of integration $y_i.$  The next step is to do conformal substitution for
each variable of integration and for each external vector of dual space of dimension $d=4-2\ve,$
\begin{eqnarray*}
y_\mu = \frac{y'_\mu}{{y'}^2}, ~~~~ {x_1}_\mu = \frac{{x'_1}_\mu}{{x'_1}^2} \Rightarrow  [y1] = \frac{[y'1']}{[y'][1']};~~~~
[y] \equiv y^2,~~~~[1] \equiv x_1^2 .   
\end{eqnarray*}
The result of this substitution is Fig.(\ref{figure20}). Omitting the prime symbols and re-arranging factors we come to Fig. (\ref{figure21}).
Now we shift all the variables of integration by some vector $x_4''$
\begin{eqnarray*}
{y_i} = {y_i}''-{x_4}'' \\
{x_1} = {x_1}'' - {x_4}'',~~{x_2} = {x_2}'' - {x_4}'',~~{x_3} = {x_3}'' - {x_4}'',
\end{eqnarray*}
in which ${y_i}$ are variables of integration in the internal vertices of Fig. (\ref{figure21}), $i=1,2,3.$ 
Omitting the prime symbols again, we obtain the expression presented in Fig. (\ref{figure22}). In this diamond we recognize the dual image of the
m.s.r.  of the box ladder diagram which is depicted in Fig. (\ref{figure23}). Indeed, this dual image  
stands on the l.h.s. of Fig. (\ref{figure24}), while the r.h.s. of  Fig. (\ref{figure24}) is read off from 
Fig. (\ref{figure22}). As we can see, this box ladder has the  index $1-\ve$ on the rungs in Fig. (\ref{figure23}).

The box ladder diagram in Fig.(\ref{figure23}) is, in some sense, equivalent to the triangle ladder in Fig.(\ref{figure16}) 
since it is represented by the same function $\phi^{(3)}$ but of the distinct arguments. This result for the box ladder diagram in Fig.(\ref{figure23})
given in Fig.(\ref{figure24}) may be generalized for an arbitrary number of rungs, 
\begin{eqnarray} \label{phi-n}
\frac{1}{s^nt^{1-\ve}} \phi^{(n)}\le\frac{p_1^2p_3^2}{st},\frac{p_2^2p_4^2}{st} \ri.
\end{eqnarray}

\section{Conclusion}

In this paper we have shown that the triangle ladder diagrams of the family  depicted in Fig. (\ref{figure16}) may be calculated explicitly
in non-integer number of dimensions $d=4-2\ve$ without making expansion in terms of $\ve$ that is a parameter of the dimensional regularization. The result has been 
achieved by a combination of analytic and dimensional regularizations. It was impossible to derive this result  
by making use of dimensional regularization only. These diagrams are calculated in terms of functions $\phi^{(n)}$ of Eq.(\ref{phi-n}). We have shown that certain family of box ladder diagrams are 
calculated in terms of these functions too.

The drawback of the result is that it cannot be used for calculation of amplitudes since the indices of rungs in the box ladders remain to be $1-\ve$ in the m.s.r. 
However, we think it is a useful result since it  may shed new light on more practical case when all the indices are equal to 1 in the m.s.r, that is for the physical box ladder
diagrams.

As to advantages of the result obtained in the present paper, we may say that it may be used 
to study the off-shell values of the physical box ladders in the theory regularized dimensionally, since the off-shell difference of our result 
of Fig. (\ref{figure24}) from the physical case  is proportional to $\ve.$ 
Among the other applications we may consider that the shifting $\ve = \ve' + \frac{1}{2},$ $\ve = \ve' + 1$ or $\ve = \ve' - 1$ would transfer the proposed generalization of
Belokurov-Usyukina loop reduction technique to three-, two- or six-dimensional scalar theories, respectively, which are regularized dimensionally. Thus, this trick may be 
generalized to an arbitrary dimension for the box ladder diagrams but with special values of indices. The exact values of indices, for which this trick can be applied, 
depend on $d.$

\subsection*{Acknowledgments}

I.K. was supported by Fondecyt (Chile) grants 1040368, 1050512, 1121030, by DIUBB (Chile) Grants 121909 GI/C-UBB  and 102609. 
His work is supported by Universidad del Bio-Bio and Ministerio de Educacion (Chile) within the project MECESUP UBB0704-PD018. He is grateful to 
Physics Faculty of Bielefeld University for accepting him as a visiting scientist, for kind hospitality and excellent conditions 
of work. Figures were drawn by means of program JaxoDraw \cite{jaxodraw}.


\begin{thebibliography}{99}




\bibitem{Broadhurst:2010ds}
D.~J.~Broadhurst and A.~I.~Davydychev, ``Exponential suppression with four legs and an infinity of loops,''
Nucl.\ Phys.\ Proc.\ Suppl.\  {\bf 205-206} (2010) 326 [arXiv:1007.0237 [hep-th]].


\bibitem{Bern:2005iz}
Z.~Bern, L.~J.~Dixon and V.~A.~Smirnov, ``Iteration of planar amplitudes in maximally supersymmetric Yang-Mills theory at three loops and beyond,''
Phys.\ Rev.\  D {\bf 72} (2005) 085001 [hep-th/0505205].




\bibitem{Smirnov} V.~A.~Smirnov, ``Evaluating Feynman Integrals,'' Springer Tracts Mod. Phys. {\bf 211}, 1 (2004) 



\bibitem{Usyukina:1992jd}
N.~I.~Usyukina and A.~I.~Davydychev, ``An Approach to the evaluation of three and four point ladder diagrams,''
Phys.\ Lett.\  B {\bf 298} (1993) 363.



\bibitem{Usyukina:1993ch}
N.~I.~Usyukina and A.~I.~Davydychev, ``Exact results for three and four point ladder diagrams  with an arbitrary number of rungs,''
Phys.\ Lett.\  B {\bf 305} (1993) 136.




\bibitem{Boos:1990rg} E.~E.~Boos and A.~I.~Davydychev, ``A Method of evaluating massive Feynman integrals,''
Theor.\ Math.\ Phys.\  {\bf 89} (1991) 1052 [Teor.\ Mat.\ Fiz.\  {\bf 89} (1991) 56].


\bibitem{Davydychev:1992xr}
A.~I.~Davydychev, ``Recursive algorithm of evaluating vertex type Feynman integrals,''
 J.\ Phys.\ A {\bf 25}, 5587 (1992).



\bibitem{Belokurov:1983km} V.~V.~Belokurov and N.~I.~Usyukina, ``Calculation Of Ladder Diagrams In Arbitrary Order,''
J.\ Phys.\ A  {\bf 16} (1983) 2811.


\bibitem{Usyukina:1983gj} N.~I.~Usyukina, ``Calculation Of Many Loop Diagrams Of Perturbation Theory,''
Theor.\ Math.\ Phys.\  {\bf 54} (1983) 78 [Teor.\ Mat.\ Fiz.\  {\bf 54} (1983) 124].


\bibitem{Usyukina:1991cp} N.~I.~Usyukina, ``Calculation of multiloop diagrams in arbitrary order,''
 Phys.\ Lett.\  B {\bf 267} (1991) 382 [Theor.\ Math.\ Phys.\  {\bf 87} (1991) 627] 
[Teor.\ Mat.\ Fiz.\  {\bf 87} (1991) 414]



\bibitem{Broadhurst:1993ib}
D.~J.~Broadhurst, ``Summation of an infinite series of ladder diagrams,''
  Phys.\ Lett.\ B {\bf 307} (1993) 132.




\bibitem{Kondrashuk:2009us}
I.~Kondrashuk and A.~Vergara, ``Transformations of triangle ladder diagrams,'' JHEP {\bf 1003} (2010) 051
[arXiv:0911.1979 [hep-th]].


\bibitem{Gonzalez:2012gu}
I.~Gonzalez, I.~Kondrashuk, 
``Belokurov-Usyukina loop reduction in non-integer dimension,'' 
Phys.\ Part.\ Nucl.\  {\bf 44} (2013) 268 [arXiv:1206.4763 [hep-th]].


\bibitem{Allendes:2012mr}
P.~Allendes, B.~Kniehl, I.~Kondrashuk, E.~A.~Notte Cuello, M.~ Rojas Medar,
``Solution to Bethe-Salpeter equation via Mellin-Barnes transform,''
Nucl.\ Phys.\ B {\bf 870} (2013) 243
[arXiv:1205.6257 [hep-th]].







\bibitem{Unique} M.~D'Eramo, L.~Peliti and G. Parisi,  ``Theoretical Predictions for Critical 
Exponents at the $\lambda$-Point of Bose Liquids,'' Lett. Nuovo Cimento 2 (1971) 878.



\bibitem{Vasiliev:1981dg} A.~N.~Vasiliev, Y.~M.~Pismak and Y.~R.~Khonkonen, ``1/N Expansion: Calculation 
Of The Exponents Eta And Nu In The Order 1/N**2 For Arbitrary Number Of Dimensions,'' 
Theor.\ Math.\ Phys.\  {\bf 47} (1981) 465
[Teor.\ Mat.\ Fiz.\  {\bf 47} (1981) 291].



\bibitem{Vasil} A.N. Vasil'ev, ``The Field Theoretic Renormalization Group in Critical Behaviour Theory and Stochastic Dynamics'', (Chapman 
\& Hall/CRC, Boca Raton, Florida, 2004)



\bibitem{Kazakov:1984bw}  D.~I.~Kazakov,  ``Analytical Methods For Multiloop Calculations: 
Two Lectures On The Method Of Uniqueness,'' JINR-E2-84-410.




\bibitem{Cvetic:2006iu} G.~Cvetic, I.~Kondrashuk, A.~Kotikov and I.~Schmidt, 
``Towards the two-loop Lcc vertex in Landau gauge,'' Int.\ J.\ Mod.\ Phys.\ A {\bf 22} (2007) 1905
[hep-th/0604112].



\bibitem{Kondrashuk:2008ec} 
I.~Kondrashuk and A.~Kotikov, ``Fourier transforms of UD integrals,''  arXiv:0802.3468 [hep-th], in {\it Analysis and Mathematical Physics}, 
Birkh\"auser Book Series Trends in Mathematics, edited by B. Gustafsson and A. Vasil'ev, Birkh\"auser, Basel, Switzerland, 2009, pp. 337-348




\bibitem{Kondrashuk:2008xq}
I.~Kondrashuk and A.~Kotikov, ``Triangle UD integrals in the position space,''  
JHEP {\bf 0808} (2008) 106 [arXiv:0803.3420 [hep-th]].


\bibitem{Allendes:2009bd} P.~Allendes, N.~Guerrero, I.~Kondrashuk and E.~A.~Notte Cuello, ``New four-dimensional integrals by Mellin-Barnes transform,'' 
J.\ Math.\ Phys.\ \textbf{51} (2010) 052304 [arXiv:0910.4805
[hep-th]]. 




\bibitem{jaxodraw}  D.~Binosi, L.~Theussl, ``JaxoDraw: A graphical user interface for drawing Feynman diagrams''
Comp. Physics Comm. {\bf 161} (2004) 76





\end{thebibliography}
\end{document}